\documentclass[fleqn,10pt]{wlscirep}
\usepackage[utf8]{inputenc}
\usepackage[T1]{fontenc}
\usepackage[revision]{revdiff}
\usepackage{hyperref}       % hyperlinks
\usepackage{url}            % simple URL typesetting
\usepackage{booktabs}       % professional-quality tables
\usepackage{amsfonts}       % blackboard math symbols
\usepackage{microtype}      % microtypography
\usepackage{lipsum}		% Can be removed after putting your text content
\usepackage{doi}

\usepackage{amsmath}
\usepackage{bbm}
\usepackage{graphicx, epsfig, nicefrac}
\usepackage{amsthm}
\usepackage{times}
\usepackage{color}
\usepackage{lineno}

\title{A Bayesian generative neural network framework for epidemic inference problems}

\author[1,*]{Indaco Biazzo}
\author[1]{Alfredo Braunstein}
\author[1]{Luca Dall'Asta}
\author[1]{Fabio Mazza}
\affil[1]{Politecnico di Torino, DISAT, Turin, 10129, Italy}

\affil[*]{indaco.biazzo@polito.it}

\begin{abstract}
The reconstruction of missing information in epidemic spreading on contact networks can be essential in the prevention and containment strategies. The identification and warning of infectious but asymptomatic individuals (i.e., contact tracing), the well-known patient-zero problem, or the inference of the infectivity values in structured populations are examples of significant epidemic inference problems.
As the number of possible epidemic cascades grows exponentially with the number of individuals involved and only an almost negligible subset of them is compatible with the observations (e.g., medical tests), epidemic inference in contact networks poses incredible computational challenges. 
We present a new generative neural networks framework that learns to generate the most probable infection cascades compatible with observations. The proposed method achieves better (in some cases, significantly better) or comparable results with existing methods in all problems considered both in synthetic and real contact networks. 
Given its generality, clear Bayesian and variational nature, the presented framework paves the way to solve fundamental inference epidemic problems with high precision in small and medium-sized real case scenarios such as the spread of infections in workplaces and hospitals.
\end{abstract}
\begin{document}

\flushbottom
\maketitle
\thispagestyle{empty}

\section{Introduction}\label{sec1}

Discrete-state stochastic compartmental models have been traditionally used to model infectious diseases \cite{Kermack1927,keeling2011modeling,Vesp_RevModPhys} and provide a simple and unified mathematical framework for a wide variety of spreading processes occurring in social and technological systems \cite{Vespignani2012}. 
The time-forward simulations of most epidemic models, even those incorporating detailed demographic and mobility data, can be efficiently performed using Monte-Carlo based sampling techniques or, at least at the meta-population level, exploiting approximation methods, such as stochastic differential equations and moment closure schemes \cite{brauer2008lecture}. 
These computational methods have been largely applied to large-scale epidemic forecasting \cite{eubank2004modelling,balcan2009multiscale,merler2011determinants,chinazzi2020effect} and containment  \cite{Ferguson2005,Halloran2008,sun2021transmission}. 
Their effectiveness crucially depends on the capacity to exploit the available information on the past behavior of the epidemic outbreak. At the meta-population level, such information,  represented by temporal series of aggregate quantities (e.g. daily number of newly infected individuals inside a reference population) can be rather easily included within traditional Bayesian computational frameworks based on Monte Carlo sampling techniques \cite{sisson2007sequential,toni2009approximate,kypraios2017tutorial}. 

\begin{figure}[ht]
\begin{center}
\includegraphics[width=1\columnwidth]{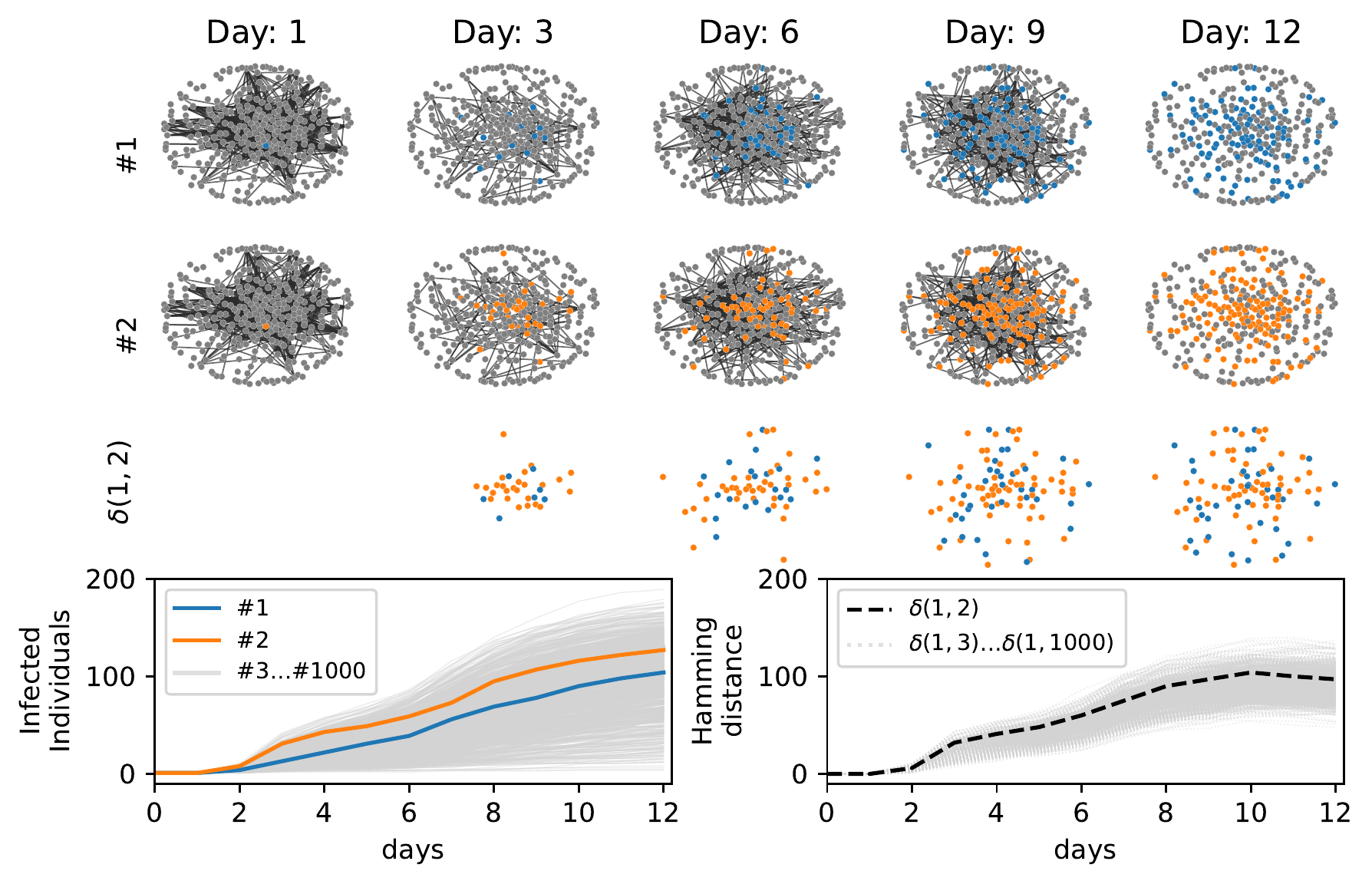}
\caption{\textbf{Simulated epidemic cascades in a hospital contact network.} One thousand epidemic cascades simulated (with the same epidemic parameters) on a real contact graph measured in a hospital\cite{Obadia2015} (detailed information about  epidemic models and contact networks are listed in the results section). The epidemics started from the same individual. Two samples (blue, and orange) of epidemic cascades are shown in the first and second rows of the figure. The third row represents the distance between them, where in this case the blue dots are the infected individuals present in the cascade 1 but not in cascade 2 and the orange ones are those present in cascade 2 but not in cascade 1. In the third row, the total number of blue and orange dots gives the Hamming distance between the two daily configurations. \textbf{Left-bottom plot.} Cumulative number of infected individuals for 1000 epidemic cascades started from the same individual. \textbf{Right-bottom plot.} Hamming distance ($\delta(1,i)$) between the cascade 1 and all the others $i\in[2,3\dots1000]$.}
\label{fig:cascade}
\end{center}
\end{figure}

The COVID-19 pandemic has motivated the interest for the large-scale adoption of epidemic surveillance techniques and digital contact tracing through smartphone applications  \cite{Ferretti2020, Wymant2021}, which could make it possible to access/use a large amount of (possibly inaccurate) individual-based observational data, traditionally available only for case studies in rather small and controlled environments \cite{mastrandrea2015contact,smieszek2016contact,sapiezynski2019interaction}. 
The availability of individual-based observational data unveils a crucial limitation of traditional Monte-Carlo based inferential techniques. While the number of possible epidemic realizations generated by a specific epidemic model on a given contact network scales exponentially with the systems size and the duration of the process, those compatible with individual-based observations are just an exponentially small fraction of them \cite{antu_identification_2015}.
It follows that inferential methods based on the direct sampling of epidemic realizations on individual-based contact networks rapidly become inefficient as the size of the outbreak increases \cite{antu_identification_2015}.
As an example, let us consider several simulated epidemic realizations (with the same initial condition, consisting of a single infected individual, and the same epidemic parameters) in a real graphs of temporal contacts between patients and staff members of a hospital \cite{Obadia2015} (see Fig.\ref{fig:cascade}). 

We define the daily configuration of the system as the daily epidemic state of the individuals (infected/not-infected) during the epidemic process. The plots in Fig.\ref{fig:cascade} indicate a fast divergence of the configurations of the simulated epidemics, even though they start from the same individual with the same epidemic parameters. Choosing, for instance, the final configuration of one epidemic cascade as the individual-based observation of the system, it is very unlikely to obtain the same configuration from a direct sampling of the epidemic model.\\
%The "blue" simulated epidemic cascade, chosen as reference, displays, during the epidemic evolution, high (hamming) distance from the all the others.

%Different strategies have been adopted to solve some of these statistical inference problems efficiently. For instance, in the case of the patient zero:
%specific topological quantity defined on the contact network (rumor centrality) \cite{shah2011rumors} or  Monte-Carlo estimator  \cite{antu_identification_2015} and message passing algorithms \cite{altarelli_bayesian_2014, PhysRevE.90.012801} have been used.

A step forward in this field is represented by the introduction of efficient algorithms for Bayesian inference based on Belief Propagation (BP) \cite{altarelli_bayesian_2014, PhysRevE.90.012801}. 
In the Bayesian inference framework, the objective is to approximately compute the posterior probability of the system, assuming the epidemic model as a prior and the individual-based observations as the evidence. 
BP-based algorithms make it possible to obtain estimates of the local marginals of the posterior distribution and, as shown in \cite{altarelli_bayesian_2014, altarelli2013optimizing,PhysRevX.4.021024}, this approach outperforms competing methods on sparse contact networks on a variety of inference problems. 
In particular, the integration of such algorithms in the framework of digital contact tracing for COVID-19 was recently shown to provide a better assessment of the individual risk and improve the mitigation impact of non-pharmaceutical interventions strategies \cite{baker2020epidemic}. 

BP-based algorithms may experience non-convergence issues, for instance in dense and very structured contact networks, a phenomenon that calls for the search of alternative inference methods which could overcome such a limitation while maintaining comparable performances on sparse networks. Here we propose to use generative neural networks, specifically autoregressive neural networks (ANN), to learn the posterior probability of an epidemic process and efficiently sample from it. In practice, the autoregressive neural network can generate realizations of the epidemic process according to the stochastic dynamical rules of the prior model but compatible with the evidence.
Deep autoregressive neural networks are used to generate samples according to a probability distribution learned from data, for instance for images  \cite{van2016pixel}, audio \cite{WaveNet},  text  \cite{bengio2003neural, keskar2019ctrl} and protein sequences \cite{trinquier2021efficient} generation tasks and, more generally, as a probability density estimator \cite{larochelle2011neural,germain2015made, uria2016neural}.  Autoregressive neural networks have recently been used to approximate the joint probability distributions of many (discrete) variables in statistical physics models \cite{PhysRevLett.122.080602}, and applied in different physical contexts \cite{PhysRevLett.124.020503,pan2021solving, PhysRevE.101.053312, Var_neural_annealing}. In this work, we show how to use a deep autoregressive neural network architecture to efficiently sample from a posterior distribution composed of a prior, given by the epidemic propagation model (even though the parameters of such model can be contextually inferred), and from an evidence given by (time-scattered) observations of the state of a subset of individuals. 
Neural networks have already been applied to epidemic forecasting \cite{app10186448, philemon2019review, CHIMMULA2020109864} but rarely to epidemic inference and reconstruction problems. Two recent preliminary works apply neural networks to epidemic inference problems: in \cite{Shah2020} the patient zero problem is tackled using graph neural networks, while a similar technique is applied to epidemic risk assessment in \cite{tomy2021estimating}.
%In \cite{Shah2020} the authors solve the patient zero problem; The performance comparison is made with only one other method, preventing a concrete assessment of the performances. In \cite{tomy2021estimating} the authors apply graph neural network to the risk assessment problems, but they do not perform any comparison with existing methods such as \cite{baker2020epidemic}. The source code of both works has not been released. 
The presented approach allows to address successfully a large class of epidemic inference problems, ranging from the patient-zero problem and individual risk assessment to the inference of the parameters of the propagation model under a unique neural network framework. We believe this to be a strong point in favour of this technique. In all such problems, the proposed autoregressive neural network architecture provides results that are at least as good as all other methods considered for comparison and outperforms them in most cases. The implementation of the algorithm and the instructions to reproduce the results are available at \cite{ann_results_github}.

\section{Methods}\label{sec11}
\subsection*{The posterior probability of the epidemic process} 
The dynamics of epidemic spreading in a contact network is commonly described by means of individual-based stochastic models in which individuals can be in a finite set of possible states, usually called epidemic compartments. For the sake of concreteness, consider the discrete-time SIR model, in which $x_i^t\in \mathcal{X} = \{S,I,R\}$ stands for the individual $i$ being at time step $t$ in the Susceptible ($S$), Infected ($I$) or Recovered ($R$) state. The infection of a susceptible individual due to a contact with an infected individual occurs with rate $\lambda$, while infected individuals recover in time with rate ${\mu}$ (heterogeneous epidemic parameters can be considered as well if necessary). In the epidemic propagation model, both the epidemic parameters and the temporal structure of the underlying contact network are assumed to be given and known, so that the individual transition probability of the corresponding Markov chain reads as follows
\begin{align}
p\left(x_i^{t+1}=S \lvert \mathbf{x}_{\partial i}^t , x_i^t \right)  & = \mathbbm{1}\left[  x^t_i = S\right] \prod_{j\in\partial i} \left(1-\lambda  \mathbbm{1}\left[x_j^{t} = I\right] \right),\\
p\left(x_i^{t+1}=R \lvert \mathbf{x}_{\partial i}^t , x_i^t \right)  & = \mathbbm{1}\left[  x^t_i = R\right] + \mathbbm{1}\left[  x^t_i = I\right]\mu,
\end{align}
and $p\left(x_i^{t+1}=I \lvert \mathbf{x}_{\partial i}^t , x_i^t \right) = 1 - p\left(x_i^{t+1}=S \lvert \mathbf{x}_{\partial i}^t , x_i^t \right) - p\left(x_i^{t+1}=R \lvert \mathbf{x}_{\partial i}^t , x_i^t \right)$. Here $\mathbbm{1}\left[  x^t_i = X\right]$ is the indicator function that is equal to $1$ when $x_t^i=X$ and zero otherwise, $\mathbf{x}_{\partial i}^t$ represents the set of individuals that are in contact with node $i$ at time $t$. 
%The configuration of all $N$ variables at time $t$ is denoted as $\mathbf{x^t}=\{x_1^t \dots x_n^t\}$, and $\underline{x}_i = \{x_i^0\dots x_i^T\}$ is the single node time trajectory.
Defining the individual epidemic-state trajectory as $\underline{x}_i = \{x_i^0,\dots, x_i^T\}$,  the probability of an epidemic realization $\underline{\mathbf{x}} = \{\underline{x}_1, \dots, \underline{x}_N\}$ between time $0$ and time $T$  is
\begin{equation}
p(\underline{\mathbf{x}})= p(x_1^0, \dots, x_N^0)\prod_{i=1}^N\prod_{t=1}^T p \left(x_i^{t} \lvert  \mathbf{x}_{\partial i}^{t-1} , x_{i}^{t-1} \right),
\end{equation}
 where $p(x_1^0, \dots, x_N^0)$ is a prior distribution on the initial state, which is usually assumed to be factorized, i.e. $p(x_1^0, \dots, x_N^0)=\prod_i p(x_i^0)$. %, where $P_i^0(x_i^0)$ and i.i.d distribution, based on a rough estimation of the ratios of $S$, $I$ and $R$ in the population at $t=0$. 

Suppose that some information about the individual states at different times is available (e.g because individuals exhibit symptoms or undergo medical tests). We will assume that this information comes in the form of a set of independent observed variables $O_r$ following known probabilistic laws $p_r(O_r\lvert x^{t_r}_{i_r})$. For example, if an individual $i_r$ has been observed in state $X_r$ at time $t_r$, we will have $p_r(O_r\lvert x^{t_r}_{i_r}) = \mathbbm{1}\left[x^{t_r}_{i_r}= X_r\right]$. False negative and positive rates in tests can be easily represented  generalizing the expression of $p_r$.

Under the assumption of independence between observations, the conditional probability of $\mathcal{O}$ given $\underline{\mathbf{x}}$ is $p(\mathcal{O}\lvert \underline{\mathbf{x}})=\prod_{r\in\mathcal{O}} p_r(O_r\lvert x^{t_r}_{i_r})$, and the posterior probability of an epidemic cascade $\underline{\mathbf{x}}$ given the observation $\mathcal{O}$ becomes
\begin{align}
p(\underline{\mathbf{x}} \lvert \mathcal{O}) & = \frac{1}{p(\mathcal{O})} p(\mathcal{O}\lvert\underline{\mathbf{x}}) p(\underline{\mathbf{x}}) \\
& = \frac{1}{Z} \prod_{i=1}^N p(x_i^0) \prod_{t=1}^T p \left(x_i^{t} \lvert \mathbf{x}_{\partial i}^{t-1} , x_i^{t-1}  \right) \prod_{r\in \mathcal{O}} p_r(O_r\lvert x^{t_r}_{i_r}) \\
& = \frac{1}{Z} \prod_{i} \Psi_i\left(\underline{x}_i, \mathbf{\underline{x}}_{\partial i}\right)
\label{eq:posterior}
\end{align}
where $\Psi_i\left(\underline{x}_i, \mathbf{\underline{x}}_{\partial i}\right) = p(x_i^0) \prod_{t=1}^T p \left(x_i^{t} \lvert  \mathbf{x}_{\partial i}^{t-1} , x_i^{t-1} \right) \prod_{r:i_r=i} p_r(O_r \lvert x^{t_r}_{i_r})$. Here, $ \mathbf{\underline{x}}_{\partial i}$ represents the set of individuals that have been in contact with $i$ at least one time during the interval $[0,T]$. The quantity  $Z=\sum_{\underline{\mathbf{x}}} {\prod_{i} \Psi_i\left(\underline{x}_i, \mathbf{\underline{x}}_{\partial i}\right)}$ is the normalization constant of the posterior probability or model evidence. Several quantities of interest can be computed from the posterior distribution.
For instance, the problem of identifying the initial source of an outbreak in a population (the so-called patient-zero problem) requires to estimate the marginal probability $p(x_i^0=I\lvert\mathcal{O})=\sum_{\underline{\mathbf{x}}} \mathbbm{1}(x_i^0=I) p(\underline{\mathbf{x}} \lvert\mathcal{O})$ 
for every individual $i$. On the other hand, if the present time is $T$, the marginal probability $p(x_i^T=I \lvert\mathcal{O})$  provides a measure of the current epidemic risk for every individual $i$. The exact computation of probability marginals requires a sum over all admissible realizations of the process, which becomes unfeasible for more than a few dozens of individuals because their number typically grows exponentially with the number of individuals. In the next subsection, we describe a method to approximately compute the marginals of the posterior distribution in Eq.\ref{eq:posterior} using Autoregressive Neural Networks (ANNs).

\subsection*{Learning the posterior probability using autoregressive neural networks}
\label{sec:learn_proc}

Given a realization $\underline{\mathbf{x}}$ of the epidemic process and a permutation $\pi=\{\pi_1,\pi_2,\dots,\pi_N\}$ of the individuals of the system, which imposes a specific ordering to the variables $\{\underline{x}_i\}$, the probability of the realization $\underline{\mathbf{x}}$ can be written as the product of conditional probabilities (chain rule) in the form
\begin{equation}
 \label{eq:cond_prob_auto}
 p(\underline{\mathbf{x}}) = \prod_{i=1}^N p(\underline{x}_i\lvert \underline{\mathbf{x}}_{<i})
\end{equation}
where $\underline{x}_i = \{x_i^0,\dots, x_i^T\}$ and $\underline{\mathbf{x}}_{<i}=\{\underline{x}_j\lvert \pi_{j}<\pi_{i}\}$ is the set of epidemic-state trajectories of individuals with label lower than $i$ according to the given permutation $\pi$. 
The distribution $p(\underline{\mathbf{x}})$ can be approximated by a trial distribution $q^{\theta} (\underline{\mathbf{x}})$ 
with the same conditional structure 
 \begin{equation}
 \label{eq:cond_prob_ann}
q^{\theta} (\underline{\mathbf{x}}) = \prod_i q_i^{\theta_i}(\underline{x}_i\lvert \underline{\mathbf{x}}_{<i}),
 \end{equation}
which can be interpreted as a (possibly deep) autoregressive neural network depending on a set of parameters $\theta = \{\theta_i\}$. From the analytical expression of the probabilistic model $p(\underline{\mathbf{x}})$, and thus that of the posterior distribution $p\left(\underline{\mathbf{x}}\lvert\mathcal{O}\right)$ defined in Eq.\ref{eq:posterior}, the operation of parameters learning can be performed using a variational approach proposed in Ref.\cite{PhysRevLett.122.080602}, in which the (reversed) Kullback-Liebler (KL) divergence 
\begin{equation}
\label{eq:dkl}
    D_{KL} \left(q^\theta \lvert\lvert p \right) = \sum_{\underline{\mathbf{x}} } q^{\theta}\left(\underline{\mathbf{x}}\right)  \ln{\frac{q^{\theta}\left(\underline{\mathbf{x}}\right)}{p\left(\underline{\mathbf{x}}\right\lvert \mathcal{O})} }
\end{equation}
is minimized with respect to the parameters $\theta$ of the trial distribution ${q^{\theta}\left(\underline{\mathbf{x}}\right)}$. The minimization of the KL divergence can be performed using standard gradient descent algorithms (see supplementary material for details).

\begin{figure}[ht]
\begin{center}
\includegraphics[width=0.95\columnwidth]{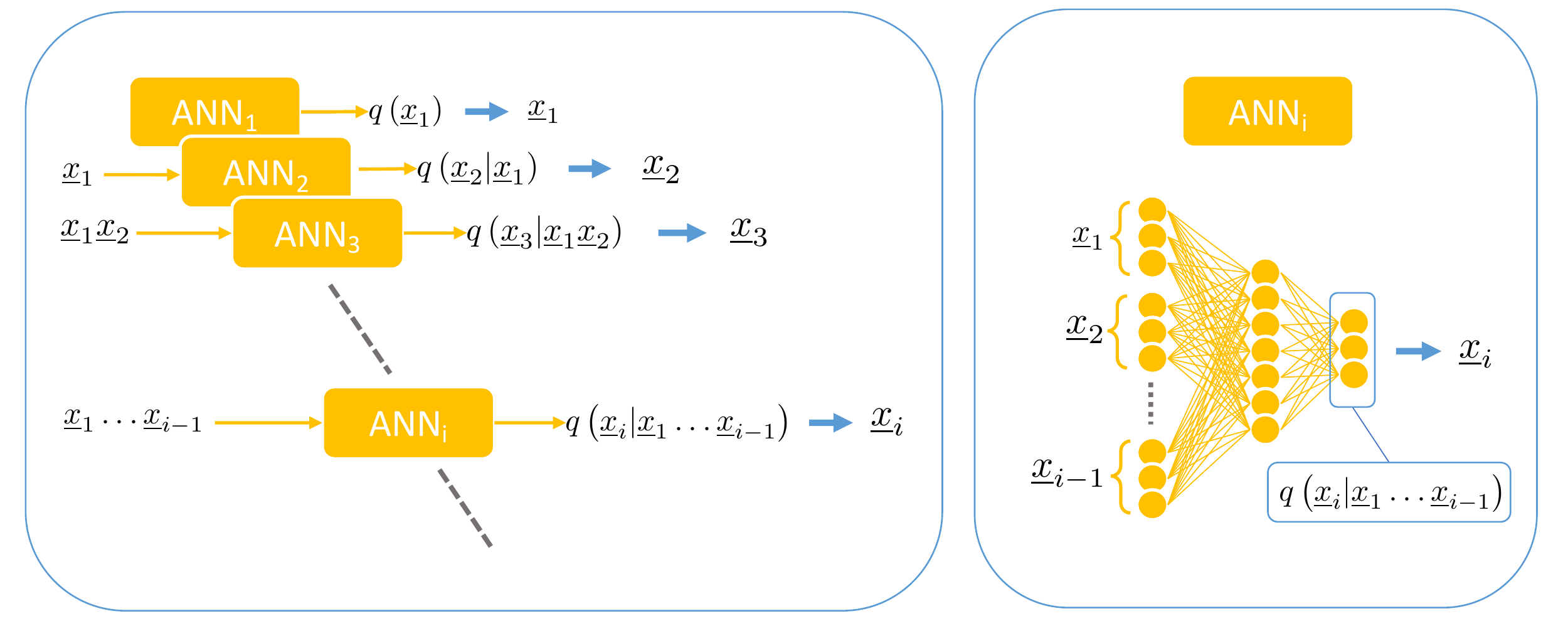}
\caption{ \textbf{Ancestral sampling of epidemic cascades.} \textbf{Left.} Ancestral sampling of epidemic cascades using artificial neural networks. For each individual $i$ there is a neural network $\text{ANN}_i$ that computes the probability $q\left(\underline{x}_i \lvert \underline{x}_{i-1}\dots \underline{x}_{1} \right)$ of its time trajectory $\underline{x}_i$ given the time trajectory of previous individuals. The time trajectory $\underline{x}_i$ is extracted from the conditional probability $q\left(\underline{x}_i \lvert \underline{x}_{i-1}\dots \underline{x}_{1} \right)$ and passed to the following neural networks. \textbf{Right.} Each neural network is composed of several fully connected layers (see supplementary material for details).}
\label{fig:net_sampling}
\end{center}
\end{figure}

The computational bottleneck of these calculation in the equations \ref{eq:dkl} and their derivatives is that the sum runs over all possible epidemic realizations, a set that grows exponentially with the size of the system. This issue is avoided by exploiting the generative power of autoregressive neural networks by training them using generated sample data through ancestral sampling. This means that the averages over the autoregressive probability distribution can be approximated as a sum over a large number of independent samples extracted from the autoregressive probability distribution $q^{\theta}$, in which the conditional structure of the autoregressive neural network allows to use the ancestral sampling procedure \cite{Goodfellow-et-al-2016}, see Fig.\ref{fig:net_sampling}.

A common way to represent the conditional probabilities in Eq.\ref{eq:cond_prob_ann} is by means of feed-forward deep neural networks with sharing schemes architectures \cite{uria2016neural,pixelCNN} to reduce the number of parameters. 
Due to the possible high variability in the dependence of $p(\underline{x}_i\lvert \underline{\mathbf{x}}_{<i})$ on $\underline{\mathbf{x}}_{<i}$ \cite{Var_neural_annealing}, instead of adopting a sharing parameters scheme we reduce the number of parameters by limiting the dependency of the conditional probability to a subset of $\underline{\mathbf{x}}_{<i}$. The subset considered is formed by all $x_j\in\mathbf{x}_{<i}$ such that $x_j$ is at most a second-order neighbor of $i$ in the graph induced by the contact network, i.e., the one in which there is an edge between two individuals if they had at least one contact during the epidemic process. The permutation order of the variables generally influences the approximation. For acyclic graphs, it is possible to define an order by which the aforementioned second-order neighbors' approximation is exact: the variables are ordered according to a spanning tree computed starting from a random node chosen as a root (see supplementary material for proof). We can imagine that the same procedure yields good approximations for sparse interacting networks, but for general interaction graphs, we are unaware of arguments for choosing an order with respect to another. In this case, random permutations of the nodes are employed.
The Kullback-Leibler divergence in Eq.\ref{eq:dkl}, could attain large (or even infinite) negative values, causing convergence issues in the parameter learning process. As illustrated in the supplementary material, this is avoided by introducing a regularization parameter, a fictitious temperature, and an annealing procedure to improve the convergence.

\subsection*{Inferring the parameters of the propagation model}
\label{sec:inf_params}
In a real case scenario, the epidemic parameters governing the propagation model are usually unknown and they should be inferred from the available data. Calling $\Lambda$ the set of these parameters (e.g. for uniform SIR models $\Lambda=\left(\lambda,\mu\right)$), the goal is to estimate them by computing the values $\Lambda^*$ that maximize the likelihood function given the set of observations $\mathcal{O}$, i.e.
\begin{align}
p(\mathcal{O} \lvert \Lambda )  & = \sum_{\underline{\mathbf{x}}} p(\mathcal{O} \lvert \underline{\mathbf{x}}) p(\underline{\mathbf{x}} \lvert \Lambda ) \\
   & = \sum_{\underline{\mathbf{x}}} \prod_i \Psi_i (\underline{x}_i)\\
   & = Z \left( \Lambda  \right).
\end{align}
The quantity $Z$ is the same normalization constant introduced in Eq.\ref{eq:posterior}, where the dependence on the parameters was dropped. Formally,
\begin{equation}
\Lambda^* = \arg \max_{ \Lambda } Z\left( \Lambda \right) = \arg \max_{ \Lambda } \log Z \left(  \Lambda  \right).
\end{equation}
Recalling that $P(\underline{\mathbf{x}}\lvert \mathcal{O})=Z^{-1}{\prod_i \Psi_i (\underline{x}_i,  \Lambda  )}$ and thanks to Gibbs' inequality we have that
\begin{align}
\log Z( \Lambda ) & = \sum_{\underline{\mathbf{x}}} p(\underline{\mathbf{x}}\lvert \mathcal{O}) \log \prod_i \Psi_i (\underline{x}_i,  \Lambda  ) - 
                 \sum_{\underline{\mathbf{x}}} p(\underline{\mathbf{x}}\lvert \mathcal{O}) \log p(\underline{\mathbf{x}}\lvert \mathcal{O}) \\
            & \geq \sum_{\underline{\mathbf{x}}} q^{\theta}(\mathbf{\underline{x}}) \log \prod_i \Psi_i (\underline{x}_i,  \Lambda  ) - 
                 \sum_{\underline{\mathbf{x}}} q^{\theta}(\mathbf{\underline{x}})\log q^{\theta}(\mathbf{\underline{x}}) \\
            & = - \left<H\right>_q + S_q
            \label{eq:gibbs_ineq}
\end{align}
where we first replaced the probability function $P(\underline{\mathbf{x}}\lvert \mathcal{O}) $ with the variational probability distribution $q^{\theta}(\mathbf{\underline{x}})$ and defined the energetic and entropic terms 
\begin{align}
\left<H\right>_q & =-\sum_{\underline{\mathbf{x}}} q^{\theta}(\mathbf{\underline{x}}) \log \prod_i \Psi_i \left(\underline{x}_i,  \Lambda\right)\\
S_q &=- \sum_{\underline{\mathbf{x}}} q^{\theta}(\mathbf{\underline{x}})\log q^{\theta}(\mathbf{\underline{x}}).
\end{align}
Since $S_q$ does not depend from $\Lambda$, minimizing $\left< H \right>_q$ with respect to parameters $\Lambda$ corresponds to maximizing $\log Z(\Lambda )$. The quantity $\left< H \right>_q$ and its derivatives w.r.t. $\Lambda$ can be computed efficiently, in an approximate way, by replacing the sum over all configurations with the average on the samples extracted by ancestral sampling from the autoregressive probability distribution $q^{\theta}$.
Therefore, we use the following heuristic procedure, inspired by the Expectation-Maximization (EM) algorithm, to infer the parameters, while minimizing the KL divergence between $q^{\theta}$ and the posterior probability $p(\underline{\mathbf{x}}\lvert \mathcal{O})$. During the learning process, two sequential steps are performed:
\begin{enumerate}
    \item Update the parameters $\{\theta_i\}$ of the autoregressive neural network to minimize the KL divergence in Eq.\ref{eq:dkl}.
    \item Update the parameters $\Lambda$ to maximize the quantity  $\left<H\right>_q$.
\end{enumerate}
These steps are repeated until the end of the learning process.

\section*{Results}

\begin{figure}[ht]
\begin{center}
\includegraphics[width=1\columnwidth]{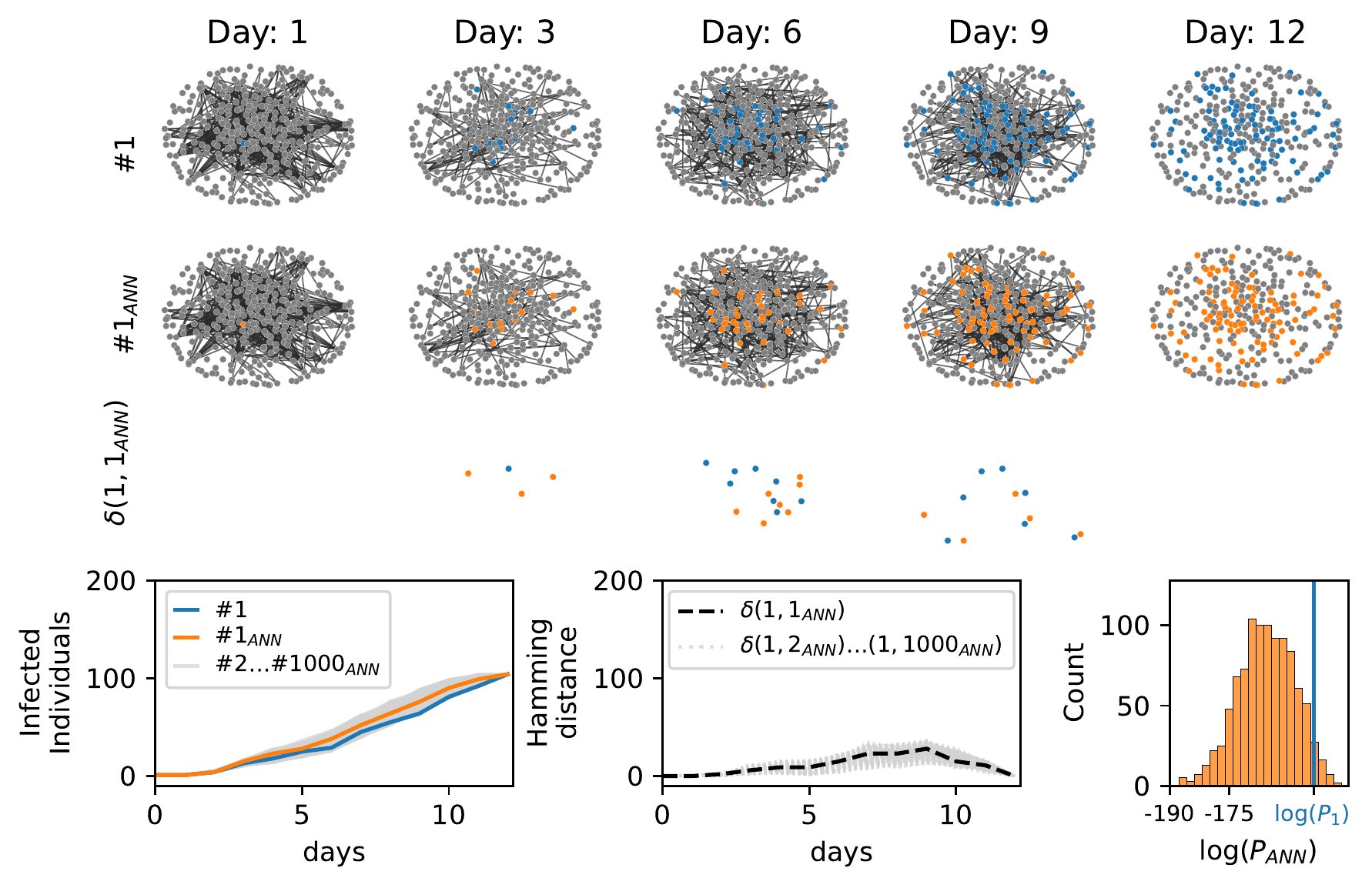}
\caption{\textbf{Epidemics cascades generated by the ANN.} Epidemic cascades generated by the ANN trained on a posterior probability composed by a prior, the epidemic model that generate the blue cascade, and the evidence, its final configuration at day 12. The contact network is a real contact graph measured in a hospital\cite{Obadia2015}. An example of the epidemic cascade generated is shown in second row ($1_{ANN}$, orange). The third row represent the hamming distance between them (see caption Fig.\ref{fig:cascade}.) \textbf{Left-bottom plot.} Cumulative number of infected individuals for epidemic cascade simulated (blue curve) and  generated by the ANN ($i_{ANN} \in [1,2\dots1000]$. \textbf{Central-bottom plot.} Hamming distance ($\delta(1,i_{ANN}$) between blue epidemic cascade and those generated by the ANN $i_{ANN}\in[2,3\dots1000]$. \textbf{Right-bottom plot.} Distribution of the values of the prior probability of the generated epidemic cascades ($P_{ANN}$). The blue vertical line is the value of the prior probability of the blue cascade ($\log(P_i$). }
\label{fig:cascade_ann}
\end{center}
\end{figure}

As a preliminary illustration of the ability of the proposed Autoregressive Neural Network (ANN) to sample epidemic realizations from a given posterior distribution, we reconsider the example in Fig.\ref{fig:cascade}, focusing on the blue epidemic cascade. We train the ANN to learn the posterior probability composed by the prior, i.e. the epidemic model that generates the blue cascade, and the evidence, i.e. its final configuration at day 12. 
The result is shown in Fig.\ref{fig:cascade_ann}. The epidemic cascades generated by the ANN have Hamming distances from the reference one that reduce to zero at day 12 (central-bottom plot) and a fraction of them have prior probabilities larger than the probability of the (blue) epidemic cascade taken as reference, right-bottom plot in Fig.\ref{fig:cascade_ann}. This example suggests that the ANN approach can generate epidemic cascades compatible with the observations and sampled according the prior epidemic model. 

In the following, we exploit the ability of the ANN to sample epidemic cascades from a posterior distribution to tackle three challenging epidemic inference problems: {\em i)} the patient-zero problem, in which the unique source of a partially observed epidemic outbreak has to be identified, {\em ii)} the risk assessment problem, in which the  epidemic risk of each individual has to be estimated from partial information during the evolution of the epidemic process, and {\em iii)} the inference of the epidemic parameters. Results are compared with those obtained using already existing methods in the field of epidemic inference. 
We also evaluate how the efficiency of the ANN algorithm depends on the size of the epidemic outbreak, measuring the number of generated epidemic samples necessary to obtain nearly optimal results. 
%Finally, we employ the ANN algorithm to perform the inference of unknown epidemic parameters. 
The comparison between different inference techniques, the Autoregressive Neural Network (ANN), a Belief Propagation based approach (SIB) (17, 33), together with the Soft Margin estimator (SM), is carried out on both random graphs and real-world contact networks. The Soft Margin (SM) estimator \cite{antu_identification_2015} is based on Monte Carlo methods in which samples are weighted according to the overlap between the observations and the generated epidemic cascade (see supplementary material for details). 
The Belief Propagation approach \cite{altarelli_bayesian_2014}, implemented in the SIB software \cite{baker2020epidemic, sib_github}  provides exact inference on acyclic contact networks and performs very effectively on sparse network structures. In the present work, we focus instead on real-world contact networks, which turns out to have relatively dense interaction patterns. The first contact network, taken from the dataset \emph{InVS13} \cite{Genois2018}, is related to a work environment ({\bf work}), while the second one was collected in a hospital ({\bf hospital}) \cite{Obadia2015}.
In both cases, the dataset used is the temporal list of contacts, respectively between $95$ and $330$ individuals, for a period of two weeks. Since the real duration $\delta^t_{i,j}$ of each contact is known, the probability of infections between individuals $i,j$ at time $t$ is computed as $\lambda^t_{i,j}=1-e^{-\gamma \delta^t_{i,j}}$, where $\gamma$ is the rate of infection.
For comparison, we also consider synthetic contact networks: a random regular graph ({\bf rrg}) with $N=100$ individuals and degree equal to $10$, and a random geometric graph ({\bf proximity}), in which $N=100$ individuals are randomly placed on a square of linear size $\sqrt{N}$. In the latter, the probability that individuals $i$ and $j$ are in contact is $e^{-d_{ij}/l}$, where  $d_{ij}$ is the distance between $i$ and $j$ and $l$ is a parameter (set to $l=10$) that controls the density of contacts. For both synthetic and empirical contact networks, epidemic processes (SIR epidemic cascades, see Material and Methods for details) with a duration of 15 days are generated. 
In the interaction graphs under study, large fluctuations in the final number of infected individuals are observed. The parameters of the epidemic model were chosen in such a way to have, on average, half of the individuals infected at the end of the epidemic propagation, in order to reduce the cases where very few or a large fraction of individuals are infected. Indeed, in these cases, the inference problems analyzed become either too trivial or very hard to solve because of lack of information. In the supplementary materials, an analysis of the robustness of the results with respect to the epidemic model parameters is shown.

\subsection*{The patient zero problem}
\label{subsec:pat_zero}
\begin{figure}[tbh]
\begin{center}
\includegraphics[width=1.05\columnwidth]{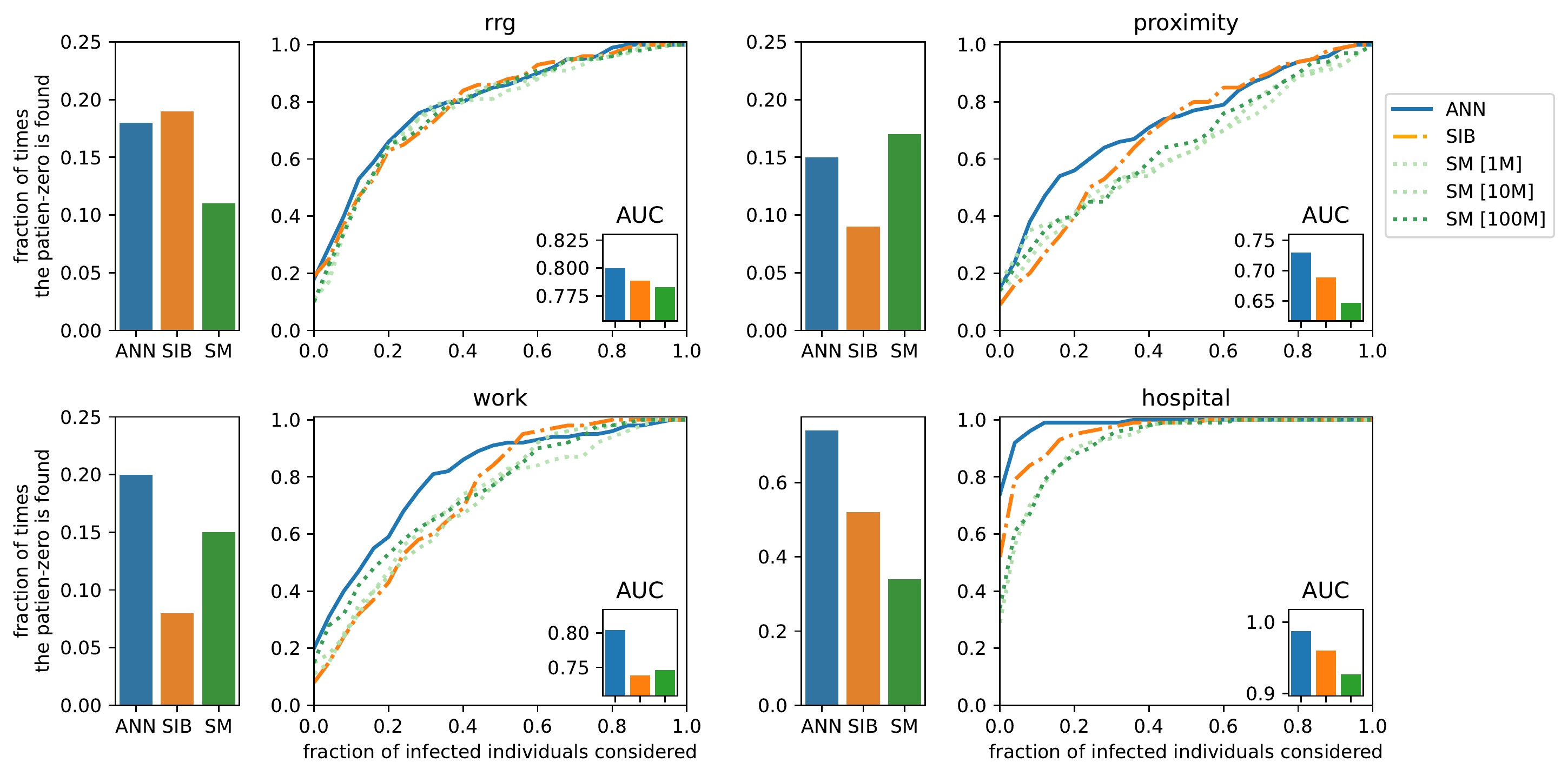}
\caption{\textbf{Results of the patient zero problem.} 
The left bar plots, for each case, represent the fraction of times, in 100 different epidemic cascades, the patient zero is correctly identified at the first position of the ranking given by the algorithms.
The right plots show the fraction of times the patient-zero is found (in 100 different epidemic cascades) in a fraction of infected or recovered individuals ranked according to the probability to be patient zero given by the three algorithms ANN, SIB, and SM (the values of the area under the curve [AUC] are shown in the insets).
For the \textbf{rrg} we consider the following epidemic parameters $\lambda=0.04$ and $\mu=0.02$ and for 
\textbf{proximity} $\lambda=0.03,\,\mu=0.02$.
The epidemic parameters for (\textbf{work}) and (\textbf{hospital}) are respectively $\gamma=10^{-3}, \,\mu=0.02$ and $\gamma=2\cdot10^{-4}, \,\mu=0.02$.}
\label{fig:p0first}
\end{center}
\end{figure}

Given the exact knowledge of the final state of the epidemics at time $T$, the patient-zero problem consists in identifying the (possibly unique) source of the epidemics. In a Bayesian framework, this problem can be tackled by computing for each individual the marginal probability of being infected at time $t=0$ given a set of observations $\mathcal{O}$. This quantity can be estimated from the posterior distribution (Eq.~\ref{eq:posterior} in the Material and Methods) with all three algorithms (ANN, SIB, and SM) considered in this work.
For each contact network ({\bf rrg, proximity, work, hospital}), we considered 100 different realizations of the epidemic model with only one patient zero. The three algorithms were used to rank infected and recovered individuals in decreasing order according to the estimated probability of being infected at time zero for each epidemic realization.
Fig.~\ref{fig:p0first} displays, for each algorithm, the fraction of times, in 100 different realizations, the patient-zero is correctly identified. The left plots show the fraction of times it is correctly identified at the first position of the infected or recovered individuals ranked according to the algorithms. The right plots show the fraction of times the patient zero is found versus the fraction of infected or recovered individuals ranked by the algorithms considered. The ANN algorithm outperforms all the other methods as indicated by the larger area under the curve (AUC) obtained in all cases considered. The improvement is also evident when analyzing the fraction of patient zero correctly identified by each algorithm (left bar plots in Fig.~\ref{fig:p0first}). For example, in the hospital case, ANN correctly identifies the patient zero in the $74\%$ of the instances, SIB in the $54\%$ and SM in the $35\%$ of them. In all cases, the ANN algorithm's performances are comparable to or better than those of the other approaches. The results on the patient zero problem reveal the ability of the ANN algorithm to efficiently generate epidemic cascades according to the posterior probability defined in Eq.~\ref{eq:posterior}. \\

\subsection*{Scaling properties with the size of the epidemic outbreak}
\begin{figure}[htb]
\begin{center}
\includegraphics[width=0.7\columnwidth]{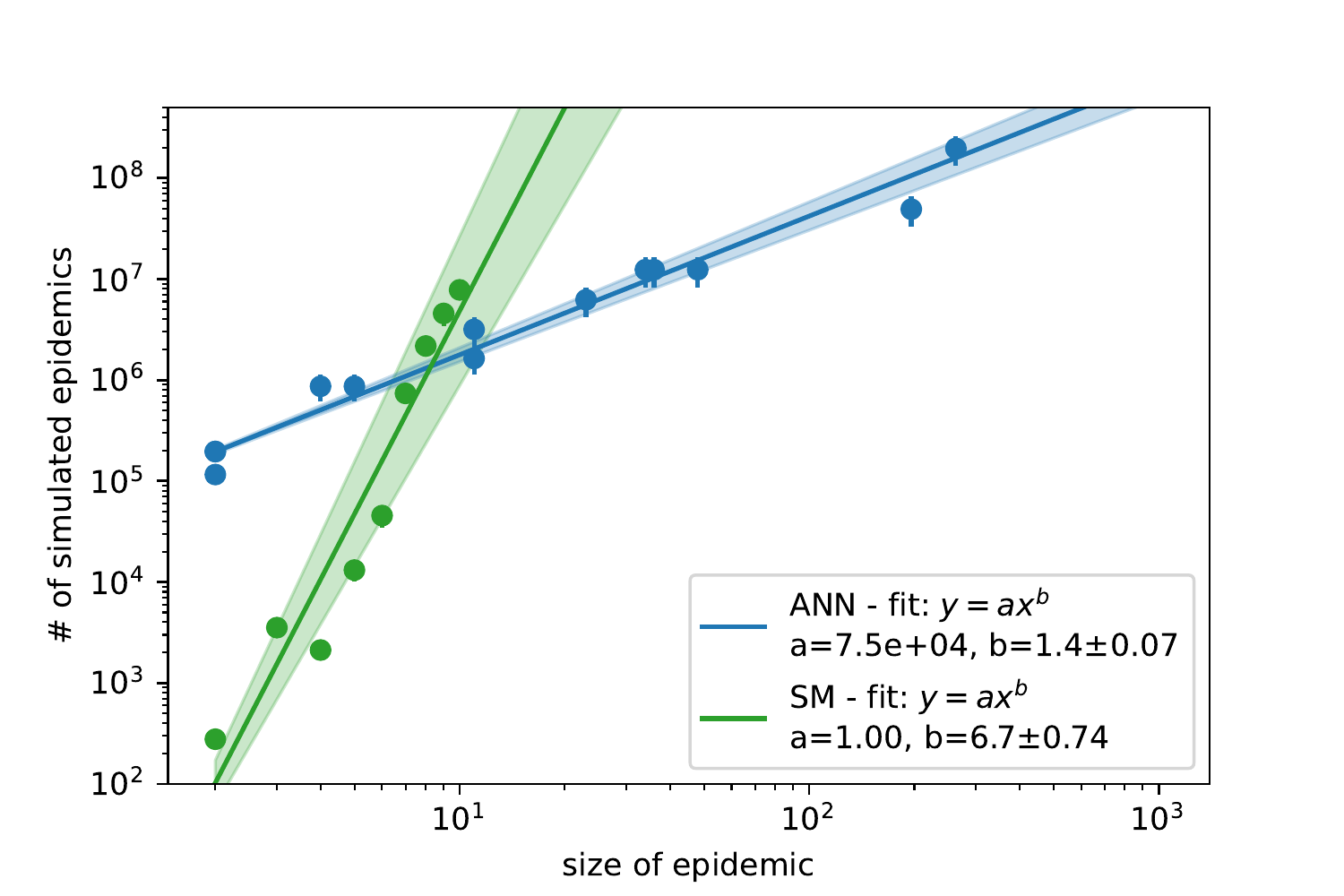}
\caption{\textbf{Scaling properties with the size of epidemic cascades.} Number of samples generated by the ANN and SM  algorithms to reach convergence. We consider the estimation of the marginal probabilities to be infected at time zero with interactions graphs given by a tree of degree and depth both equal to 6 and spanning 15 days of duration. The epidemic cascade are generated with $\mu=0$ and different values of $\lambda$ ($\lambda \in [0.1, 0.6]$). For the ANN algorithm, we consider the number of samples generated during the learning process, that is $10^3$ samples for each annealing steps (see supplementary material for details). For each instance, we run the annealing process with $2^n$ number of steps with $n\in\{5,6,\dots,18\}$. Each point is a single instance, if the algorithm converges between $2^{n-1}$ and $2^n$ steps, the number of samples reported in the plot is the number of steps $\frac{2^n + 2^{n-1}}{2} \pm \frac{2^n - 2^{n-1}}{2}$ multiplied by $10^3$ samples extracted at each step.
For the SM algorithm, %the plot displays the average number of simulated epidemics. %Since strong fluctuations are observed in the number of simulated epidemics necessary to obtain convergence to good values of the marginals,
each point in the plot is the average number of simulated epidemics necessary to reach convergence to a good estimate of the marginals (worst $10\%$ results were discarded). No point is reported when more than ten infected individuals are observed, because more than $10\%$ of the instances did not converge within $2\cdot10^8$ simulated epidemics. }
\label{fig:convergence}
\end{center}
\end{figure}

From the results presented in the previous subsection, Autoregressive Neural Networks seems to be very effective in tackling classical epidemic inference problems, particularly on dense contact networks, where the performances of BP-based methods are expected to decrease. It is, however, critical to check how the convergence property of the learning processes scales with the size of the epidemic outbreak. For this analysis, we consider the patient zero problem on a tree contact network with a unique epidemic source and where the state of the system at the final time $T$ is fully observed. With this choice, we ensure that the probability marginals computed by the SIB algorithm are exact; hence they can be taken as a reference to compare the performances of the other algorithms. 
The ANN algorithm with a second-order neighbors approximation is  exact when the interaction graph is acyclic (see supplementary material for details), assuming that the architecture of the neural networks used is sufficiently expressive to capture the complexity of posterior probability. On the other hand, since the SM algorithm is based on a Monte Carlo technique, it can give estimates of marginal probabilities with arbitrary accuracy when a sufficiently large sample of epidemic cascades is generated.\\
In the case of complete observation of the final state, the larger the epidemic size (i.e., the total number of infected individuals at time $T$), the larger is the number of epidemic cascades that are compatible with the observation. For instance, in an epidemic realization of duration $T$ time steps in which $n_I$ individuals are tested infected and $N-n_I$ tested susceptible at time $T$, the number of epidemic configurations compatible with the observations scales as $T^{n_I}$. Both ANN and SM rely on sampling procedures, so their performances could suffer from convergence issues when the epidemic size ($n_I$) increases.
We compute the total number of samples generated by the ANN during the learning process and the number of samples of epidemic cascades generated by SM in the Monte Carlo procedure. In both cases, we assume that convergence is reached when $\sum_i \lvert P_{\rm algo}(x_i^0=I\lvert \mathcal{O}) - P_{\rm sib}(x_i^0=I\lvert \mathcal{O}) \lvert < 0.1$ with ${\rm algo} \in \{ANN, SM\}$, where $P_{{\rm algo}}(x_i^0=I\lvert \mathcal{O})$ is the estimated marginal probability that individual $i$ is infected at the initial time according to each method.  
The results on the scaling properties of the ANN and SM as a function of the epidemic size on a tree contact network with degree and depth both equal to 6 ({\bf tree}) are shown in Figure \ref{fig:convergence}. Here we set the duration of the epidemic cascades to $T=15$ days. The ANN algorithm has a quasi-linear dependence with the epidemic size; conversely, the SM algorithm exhibits a very sharp increase in the number of simulations necessary for good estimates of the marginals, and already for epidemic sizes of order ten individuals, good estimates are difficult to obtain.

\subsection*{Epidemic risk assessment}
The risk assessment problem consists in finding the individuals who have the highest probability of being infected at a specific time given a partial observation $\mathcal{O}$. In particular, we consider here a realization of the SIR model with $\mu=0$ (i.e. only the states $S$ and $I$ are available) where half of the infected individuals are observed with certainty at the final time $T$. 
The results of the risk-assessment analysis obtained by means of the ANN algorithm are compared with those provided by the SIB algorithm and two other methods recently proposed in \cite{baker2020epidemic}. The Simple-Mean-Field (SMF) algorithm is based on a mean-field description of the epidemic process in which information about the observed individuals is heuristically included. The Contact Tracing (CT) algorithm computes the individual risks according to the number of contacts with observed infected individuals in the last $\tau=5$ time steps (days).

A measure of the ability to correctly identify the unobserved infected individuals at the final time $T$ is represented by the area under the Receiving Operating Characteristic (ROC) curve. This quantity, averaged over 100 instances of the epidemic realizations, is reported, for the methods considered above, in Table~\ref{table:auc_infect}, for different contact networks ({\bf rrg}, {\bf proximity} and  {\bf work}). All algorithms perform similarly on random graphs, whereas ANN and SIB outperform the other two methods in the case of the {\bf work} contact network.

\begin{table}[tb]
\begin{center}
\begin{tabular}{c|c|c|c|c|}
\cline{2-5} \cline{3-5} \cline{4-5} \cline{5-5} 
\multicolumn{1}{c|}{} & {\bf rrg} 1 src & {\bf rrg} 2 src & {\bf proximity} 1 src & {\bf work} 1 src\tabularnewline
\hline 
\hline 
ANN & $0.710\pm0.010$ & $0.670 \pm 0.009$ &$ 0.734 \pm 0.010$ & $0.889 \pm 0.005$  \tabularnewline
\hline 
SIB & $0.710 \pm 0.010$ & $0.671 \pm 0.009$ &$0.732 \pm 0.010$ & $0.886 \pm 0.005$  \tabularnewline
\hline 
SMF & $0.704 \pm 0.010$ &  $0.671 \pm 0.009$ &$0.724 \pm 0.009$ & $0.796 \pm 0.007$ \tabularnewline
\hline 
CT &  $0.685 \pm 0.009$& $0.659 \pm 0.008$ & $0.711 \pm 0.008$ & $0.790 \pm 0.006$\tabularnewline
\hline 
\end{tabular}
\caption{\label{table:auc_infect} \textbf{Epidemic risk assessment results.} 
 Area under the Receiving Operating Characteristic (ROC) curves for the risk assessment problem on random regular graphs ({\bf rrg}) with $1$ and $2$ sources, on the {\bf proximity} random graphs and {\bf work} real-world contact network. The results are averaged over $100$ different epidemic cascades generated with the same epidemic parameters. For each case, the ROC curve for the classification of the unobserved infected individuals at the final time is computed. In the {\bf rrg} case, the epidemic parameters are $\lambda_{{rrg}}^{(1)}=0.035$ for the single source and $\lambda_{{rrg}}^{(2)}=0.03$ for the double source case. For the {\bf proximity} random graphs, $\lambda_{{prox}}=0.03$. In the case of the \textbf{work} network, the model has rate of infection $\gamma_{\rm work}=10^{-3}$. 
 %The error shown is the standard deviation of the values divided by the square root of the number of instances.
 }
\end{center}
\end{table}

\subsection*{Epidemic Parameters inference}
\begin{figure}[ht]
\begin{center}
\includegraphics[width=1\columnwidth]{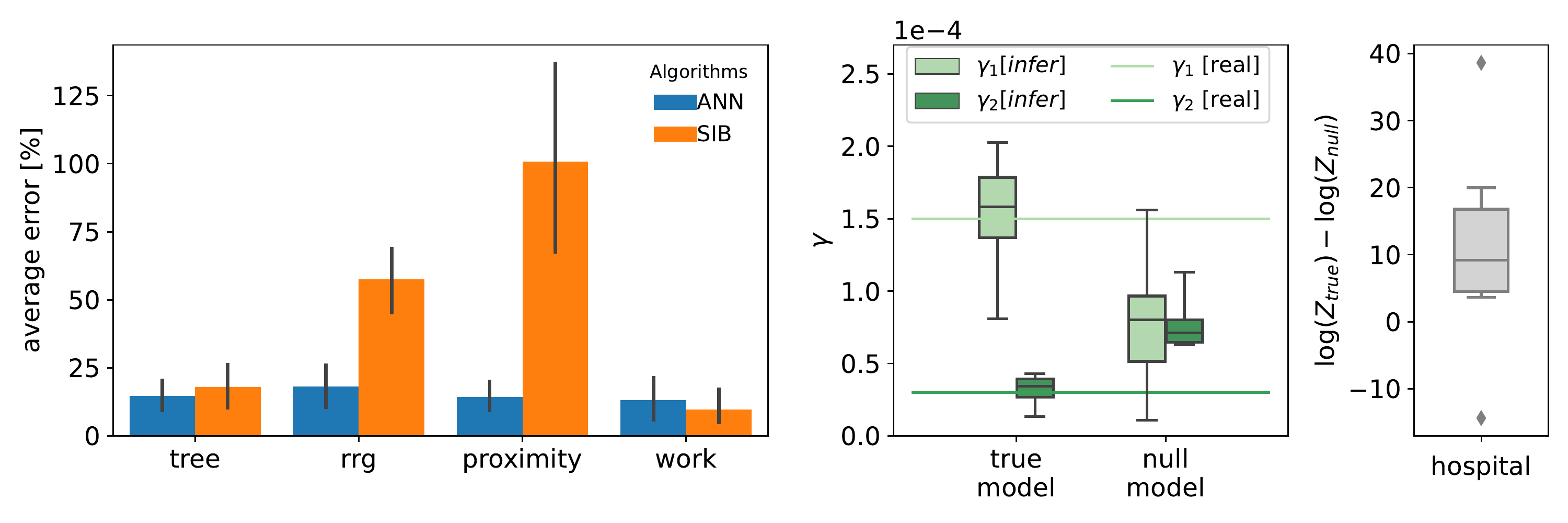}
\caption{\textbf{Inference of epidemic infectiousness parameters.} \textbf{Left plot.} Average relative error in the inference of the infectiousness parameters over ten epidemic cascade per interaction graph. On \textbf{tree}, \textbf{rrg} and \textbf{proximity} networks, the discrete-time SIR model has infection probability respectively equal to $\lambda_{\rm tree} = 0.35,\lambda_{\rm rrg} = 0.04,\lambda_{\rm proximity} = 0.03$. The  \textbf{work} case has rate of infection $\gamma_{\rm work}=10^{-3}$. The initial conditions for the parameter learning process were set to $\lambda_{\rm init}=0.5$ for \textbf{tree}, $\lambda_{\rm init}=0.1$ for \textbf{RRG} and \textbf{proximity} networks and to $\gamma_{\rm init}=10^{-2}$ for the \textbf{work} network. \textbf{Central plot.} Box plot for the case of two classes of individuals with different rate of infection $\gamma_1, \gamma_2$ inferred by the ANN. We consider two inference model where the population is divided according the propagation model (\emph{true} model) and randomly (\emph{null} model), see the text for details. The \emph{true} model is able to correctly infer the parameters with only ten different epidemic cascade. \textbf{Right plot.} Box blot of the log-likelihood difference between the \emph{true} and \emph{null} model.}
\label{fig:learn_params}
\end{center}
\end{figure}

The parameters $\Lambda$ governing the epidemic process can be simultaneously inferred during the learning process of the ANN algorithm using a heuristic method inspired by Expectation Maximization (see Materials and Methods).
Other iterative algorithms, such as SIB, can incorporate such a parameter likelihood climbing step during their convergence \cite{braunstein_network_2019}. A comparison between the performances of the two algorithms in learning the infectiousness parameter governing the spreading process on different contact networks ({\bf tree}, {\bf rrg}, {\bf proximity} and {\bf work}) is displayed in Fig.\ref{fig:learn_params} (left plot), in which we adopt the same setting of the patient-zero problem where the states of all individuals are known at the final time $T$. 
The ANN algorithm largely outperforms SIB in \textbf{rrg} and \textbf{proximity} graphs, obtaining comparable results for the \textbf{tree} and \textbf{work} instances. We also test the performance of parameters inference in a more challenging scenario where the population is split into two classes, with two different rates of infections $\gamma_1, \gamma_2$ (which could correspond, for instance, to a simplified scenario of vaccinated/not-vaccinated individuals). The states of all individuals at final time $T=14$ are observed for ten epidemic cascades on the \textbf{hospital} contact network. Then we infer the parameters with two different epidemic models: in the first one, the population is correctly divided (we call this the \emph{true} model); in the second, we split the population randomly (\emph{null} model).
The goal is to verify whether the \emph{true} model has a larger likelihood than the \emph{null} model, that is it can better explain the observations. In the central plot of Fig.\ref{fig:learn_params}, we observe how well the \emph{true} model can infer the correct values of the infections rate of the two sub-populations. As expected, the two values of $\gamma$ inferred with the \emph{null} model are similar to each other but different from the correct ones. From the rightmost plot of Fig.\ref{fig:learn_params}, we observe that the log-likelihood of the \emph{true} model is much larger of the one of the \emph{null} model, indicating the former better explains the observations. This example shows how the ANN approach can therefore be used to select the epidemic model that best explains the observations based on the estimate of their log-likelihood.

\section{Discussion}\label{sec12}

This work shows how significant individual-based epidemiological inference problems defined on contact networks can be successfully addressed using autoregressive neural networks. In problems such as patient zero detection and epidemic risk assessment, the proposed method exploits the generative power of autoregressive neural networks to learn to generate epidemic realizations that are sampled according to the epidemic model and, simultaneously, are compatible with the observations. When the model parameters are unknown, it can also infer them during the learning process. The approach is flexible enough to be easily applied to other epidemic inference problems and with different propagation models. The proposed architectures for the autoregressive networks significantly reduce the number of necessary parameters with respect to vanilla implementation. Moreover, convergence properties are improved by means of a regularization method that exploits the introduction of a  fictitious temperature and an associated annealing process. 

%The approach is 

According to the results obtained on three different problems (patient zero, risk assessment, and parameters inference) on both synthetic and real contact networks, the proposed method equals the currently best methods in the literature on epidemic inference, outperforming them in several cases. In particular, the ANN approach is computationally less demanding than standard Monte Carlo methods, as shown in fig.\ref{fig:convergence}, where the number of samples generated to reach  convergence scale almost linearly with the epidemic size. 
More efficient algorithms based on message-passing methods, like SIB, might experience convergence issues on dense contact networks like those measured in a hospital and a work office, and in these cases ANN provides significantly better results, as figure \ref{fig:p0first} shows. The framework proposed combines the high expressiveness of the neural networks to represent complex discrete variable probability distributions and the robustness of the gradient descent methods to train them. 
Moreover, the technique is a variational approach based on sampling of the distribution, which allows to compute an approximation of the log-likelihood, enabling model selection as shown in fig.\ref{fig:learn_params}.
On the other hand, like most neural networks approaches, the proposed framework suffers from some degree of arbitrariness in the choice of the neural network architecture and, consequently, the number of parameters. Moreover, in our approach, we have to pick an ordering of the variables, which could influence the quality of the approximations. 
In the supplementary material, an optimal order is shown to exist for acyclic contact networks, but it is unclear how to generalize this result  on different systems.  
These limitations are the subject of very active research in different domains where neural networks find application; within our method, the fact that an approximation of the log-likelihood is computed could help to find and test schemes and architectures suitable for each particular case. 

Although showing suitable scaling properties with the system's size, our framework reasonably needs improved architectural schemes and learning processes to be applied in epidemic inference problems regarding more than few thousand individuals. 
Work is in progress in this direction, possibly guided by symmetries and regularity of the prior epidemic models.

For all these reasons, the method seems very promising for epidemic inference problems defined in small communities such as hospitals, workplaces, schools, and cruises, where contact data could be available. In such contexts, it could detect the source of an outbreak, measure the risk of individuals being infected to improve contact tracing, or estimate the channels of contagion and the infectivity of classes of people, thanks to the possibility of inferring the propagation parameters.

\section*{Acknowledgments}

This work has been supported by the SmartData@PoliTO center on Big Data and Data Science. We acknowledge computational resources by HPC@POLITO (\href{http://www.hpc.polito.it}{http://www.hpc.polito.it}). 

\section*{Author contributions}
I.B. conceived the original idea, I.B and F.M. implemented the algorithms, I.B., A.B., L.D.A., F.M. contributed to the development of the project, the analysis of the results, and wrote the manuscript.

\section*{Declarations}
The authors declare that they have no competing interests. 

\section*{Data availability}
All data and code needed to evaluate the conclusions are released on GitHub:  \href{https://github.com/ocadni/annfore-results}{annfore-results}.
DOI: \href{10.5281/zenodo.6794183}{10.5281/zenodo.6794183}

\bibliography{references}

\end{document}


%\baselineskip24pt


\maketitle 

\section{Learning procedures and regularization}

The learning procedure involves the minimization of the Kullback-Leibler divergence, eq.9 in the main text, between the posterior probability and the variational autoregressive neural networks. More precisely, the expression of the posterior can be factorized as a product over nodes (individuals) and times, $p \left( \underline{\mathbf{x}} | \mathcal{O} \right) \propto \prod_i \Psi_i(\left(\underline{x}_i, \mathbf{\underline{x}}_{\partial i}\right)) = \prod_{i,t} \psi_i^t \left(x_i^t, \mathbf{x}^{t-1}_{\partial i}\right)$ where
\begin{equation}    
\psi_i^t \left(x_i^t, \mathbf{x^{t-1}_{\partial i}}\right) = 
\begin{cases}     
p_i^t \left(x_i^{(t)} | \mathbf{x}^{t-1}_{\partial i} \right)  \prod_{(i_r,t_r) = (i,t)} p_r(O_r^{t_r} \mid x_i^t) & t>0\\
p_i(x_i^0) \prod_{(i_r,t_r) = (i,t)} p_r(O_r^{t_r} \mid x_i^t)   & t=0\\
\end{cases}
\end{equation}
Since the factor $\psi_i^t$ can be zero for some epidemic configuration $\underline{\mathbf{x}}$, for example when they are forbidden by dynamical constraints, the gradient of the KL divergence, Eq.12 in the main text, can diverge preventing the application of gradient descent algorithms. In these cases, the diverging term $\log \psi^t_i = -\infty$ is replaced with a regularization term $\log\epsilon$ with $\epsilon\ll 1$. 
The gradient of the KL divergence with respect to the parameters of the trial distribution reads 
\begin{align}
    & \nabla_{\theta} D_{KL}(q \lvert\lvert p) = \nabla_{\theta} \sum_{\underline{x}} q_{\theta}\left( \underline{\mathbf{x}} \right) \left[-\log p \left( \underline{\mathbf{x}} \lvert \mathcal{O} \right) + \log q_{\theta}\left( \underline{\mathbf{x}} \right)
    \right]\\
    & = \sum_{\underline{x}} \nabla_{\theta} q_{\theta}\left( \underline{\mathbf{x}} \right)  \left[-\log \left( \prod_{i} \Psi_i\left(\underline{x}_i, \mathbf{\underline{x}}_{\partial i}\right) \right)+  \log q_{\theta} \left( \underline{\mathbf{x}} \right) \right] 
    + \sum_{\underline{x}} q_{\theta}\left( \underline{\mathbf{x}} \right)  \nabla_{\theta} \log q_{\theta} \left( \underline{\mathbf{x}} \right) \\
    & = \sum_{\underline{x}} q_{\theta}\left( \underline{\mathbf{x}} \right)  \left[-\log \left( \prod_{i} \Psi_i\left(\underline{x}_i, \mathbf{\underline{x}}_{\partial i}\right) \right)+ \log q_{\theta}\left( \underline{\mathbf{x}} \right)  \right] \nabla_{\theta} \log q_{\theta} \left( \underline{\mathbf{x}} \right), 
    \label{eq:dkl_der}
\end{align}
where we used that $\sum_{\underline{x}} q_{\theta}\left( \underline{\mathbf{x}} \right)  \nabla_{\theta} \log q_{\theta} \left( \underline{\mathbf{x}} \right) = \sum_{\underline{x}} \nabla_{\theta} q_{\theta}( \underline{\mathbf{x}}) =0$ and $\log(Z)\sum_{\underline{x}} \nabla_{\theta} q_{\theta}( \underline{\mathbf{x}})=0$ due to the normalization $\sum_{\underline{x}}q_{\theta}( \underline{\mathbf{x}})=1$.
The presence of large negatives values in the derivatives (e.g. due to $\log \epsilon$ terms) reduces the ability of gradient descent algorithms to explore all configurations compatible with the constraints. To overcome this issue, an annealing procedure is adopted, in which a fictitious inverse temperature $\beta$ is introduced in the computation the gradient of the KL divergence,
\begin{equation}
    \nabla_{\theta} D_{KL}^{\beta}(q|p)  = \sum_{\underline{x}} q_{\theta} \left[-\beta \log \left( \prod_{i} \Psi_i\left(\underline{x}_i, \mathbf{\underline{x}}_{\partial i}\right) \right) + \log q_{\theta} \right] \nabla_{\theta} \log q_{\theta}.
    \label{eq:dkl_der_beta}
\end{equation}
The minimization procedure starts with $\beta=0$, where all the configurations are allowed with uniform probability, then the parameter $\beta$ is slowly increased until it reaches $\beta=1$, at which the original expression of the loss function is recovered.

\section{Neural network architecture}\label{sec_archit}
In the proposed approach, each conditional probability function $p_i (\underline{x}_i|\underline{\mathbf{x}}_{<i})$ has to be approximated with a neural network $q^{\theta_i}_i\left(\underline{x}_i|\underline{\mathbf{x}}_{<i} \right)$. For monotonic models such as the SIR epidemic model, the state-space representation of temporal trajectories as a sequence $\underline{x}_i\in \mathcal{X}^{T+1}$ of $T+1$ individual states is redundant, and it turns out to be more efficient to represent them by the time instant $t_i^{I}\in(0, T+1)$ at which individual $i$ becomes infected and time instant $t_i^{R}\in (0, T+1)$ at which it recovers, the time $T+1$ corresponding to the case of no infection or recovery occurred for $i$ in the interval $[0,T]$. More precisely, the case $\left(t_i^{I}=T+1,\,t_i^{R}=T+1\right)$ corresponds to individual $i$ being susceptible at every time of the epidemic process, and $\left(t_i^{I} \neq T+1,\,t_i^{R}=T+1\right)$ corresponds to individual $i$ being infected but not recovering  in the time interval $[0,T]$.

This parametrization reduces the state space that is explored by our method, decreasing the number of incorrect time trajectories generated (for instances, time trajectory with transitions from state $I$ to $S$).

In order to compute the probability of a trajectory of a single individual, we employ two neural networks, one for the time of infection, and one for the recovery time.
Each of these networks is given as input a subset of the time trajectories of individuals with lower sorting index $\pi_i$ (see next section), and the one of the recovery time is also given as input the infection time of the same individual. The input of the networks are the infection and recover times of previous individuals, encoded in \emph{one hot} scheme.
Therefore, the (conditional) probability distributions represented by the networks are, for each individual $i$, $q_i^{\theta_{i,I}} \left( t_i^{I} \mid \left\{  t_j^{I}, t_j^{R} \right\}_{\pi_j < \pi_i}\right)$ for the infection times, and $q_i ^{\theta_{i,R}} \left(t_i^{R} \mid \left\{  t_j^{I}, t_j^{R} \right\}_{\pi_j < \pi_i}, t_i^{I}\right)$ for the recovery times, where $\theta_{i,I}$ and $\theta_{i,R}$ are the weights of each network. 

In our implementation, each neural network is a multi-layer perceptron (MLP) composed of three hidden layers plus one output layer; each layer is fully connected, and can be written as $ \underline{L}_{k+1} = \sigma \left( \mathbf{W_k}  \underline{L}_k + \underline{b}_k \right) $ where $ \underline{L}_k$ is the input vector (output of layer $k$), $\mathbf{W_k} \in \theta$ is the matrix of the weights, $\underline{b}_k \in \theta$ is the bias vector, and $\sigma $ is the activation function, which is \emph{Relu} for the hidden internal layers and \emph{Softmax} for the last layer.
The width of each layer (the number of neurons) varies linearly from the input size of each network to the output size. 

The size of the input and outputs of the neural networks used for each individual depends on the observations and the contact graph in the following way: a) the input size depends on the number of individuals with lower index that are considered (depending on the approximation made, see section \ref{sec_neigh_approx}) and
b) the observations made on an individual restrict the phase space of the possible instants of infection (for example, if we observe individual $i$ is in the infected state at time $t_O$, this means that his/her infection time $t^I_i \leq t_O$ and the recovery time $t^R_i > t_O$).
This last condition affects the size of the output of each network, but also the input size of the networks for the individuals with higher index, whose network depends on it: for example, if individual $i$ is observed in state $S$ at the final time ($T$) with certainty, this implies that the infection and recovery times are fixed, $t_i^I = T+1 $ and $t_i^R = T+1 $, and the network will always give the same value for them. Thus, in this case, for all individuals $j$ with $\pi_j > \pi_i$, the dependency on $i$ will effectively disappear.

\subsection{Graph approximations}\label{sec_neigh_approx}
We now discuss the approximation to the dependency of the conditional probabilities for the individuals' trajectories. The probability of a whole epidemic cascade can be written in the following autoregressive expression:
\begin{equation}
 %\label{eq:cond_prob_ann}
q^{\theta} (\underline{\mathbf{x}}) = \prod_i q_i^{\theta_i}\left(\underline{x}_i| \underline{\mathbf{x}}_{<i} \right),
 \end{equation}
The dependencies in the factors $q_i^{\theta_i}(\underline{x}_i| \underline{\mathbf{x}}_{<i})$ can be reduced, in the case of acyclic contact networks, to only those corresponding to the next-nearest neighbors with lower index (we relegate the derivation to section \ref{sec_acyclic_graph}), but in the general case of contact networks with cycles it is just an approximation, which we employ for all contact networks analyzed in the present work. 

In order to check this approximation, we have run several tests on the \textbf{patient zero} problem (see main text for the definition), applied to the proximity random contact graphs. For this purpose, we train the ANN either (a) considering the full dependencies on all nodes with lower permutation index (we call this version \emph{full graph}),
(b) consider only the dependency on the first and second neighbors (called \emph{next nearest neighbors} approximation) with lower permutation index, which is the approximation used in all the rest of this work, (c) considering only the first neighbors (\emph{nearest neighbors}) with lower permutation index, or (d) ignoring the dependency on the rest of the graph (called \emph{mean field} approximation).

\begin{figure}
\centering
\includegraphics[width=0.7\textwidth]{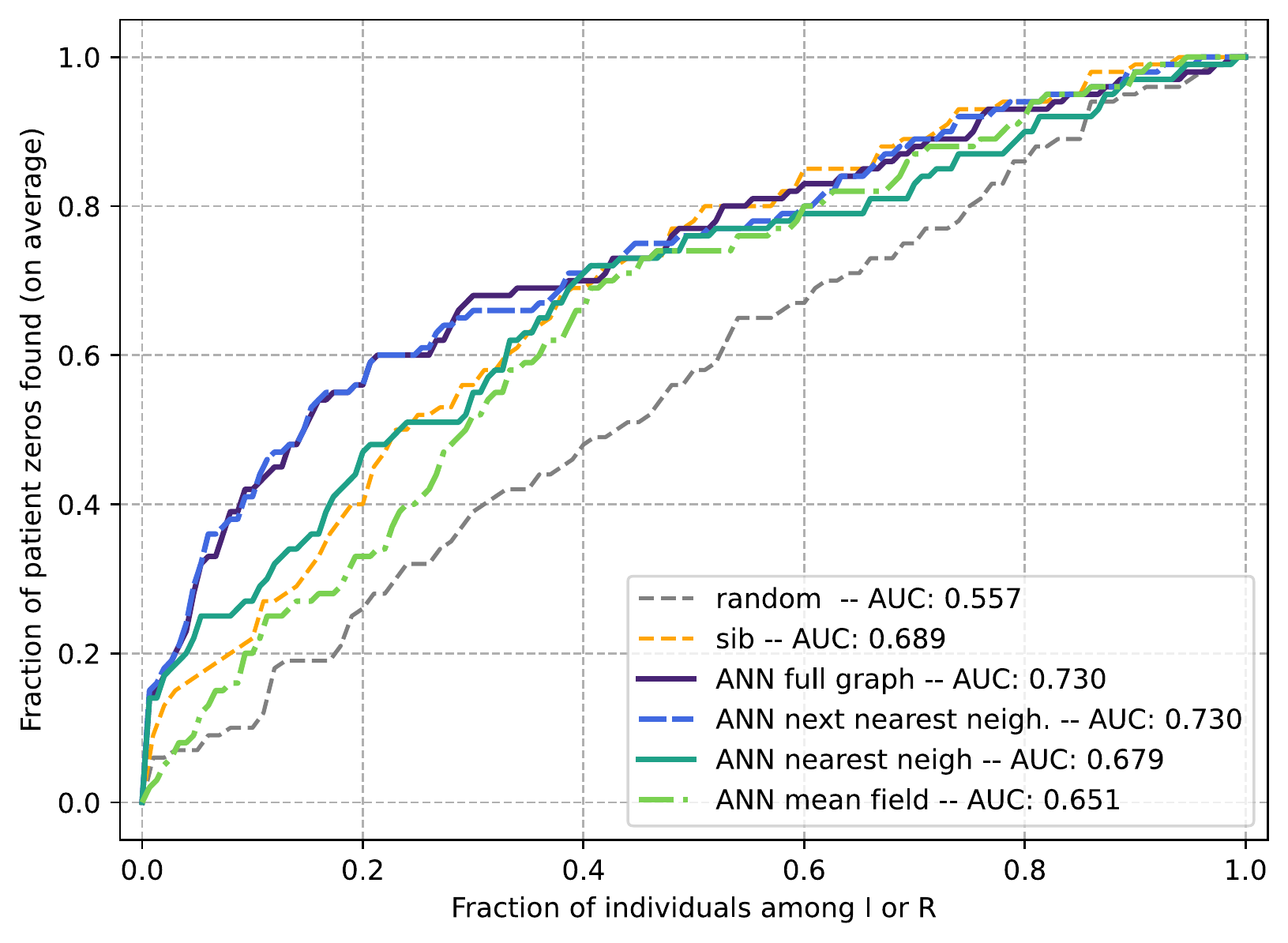}
\caption{Accuracy in finding the patient zero with different approximations of the dependencies between individuals, compared with sib (Belief Propagation) method and random guessing of the source among the infected. The "full graph" case is when no approximation on the dependencies is made. The average is made over 100 epidemic cascade on contacts network generated from a \textbf{proximity} random graph (see main text) with $N=100$ individuals. The epidemic parameters used fro the generation are $\lambda = 0.03$ and $\mu = 0.02$, with $T=15$ time instants. The legend also shows the Area under each curve (AUC).}
\label{fig:pzeroaccuANN}
\end{figure}

\begin{figure}
\centering
\includegraphics[width=0.6\textwidth]{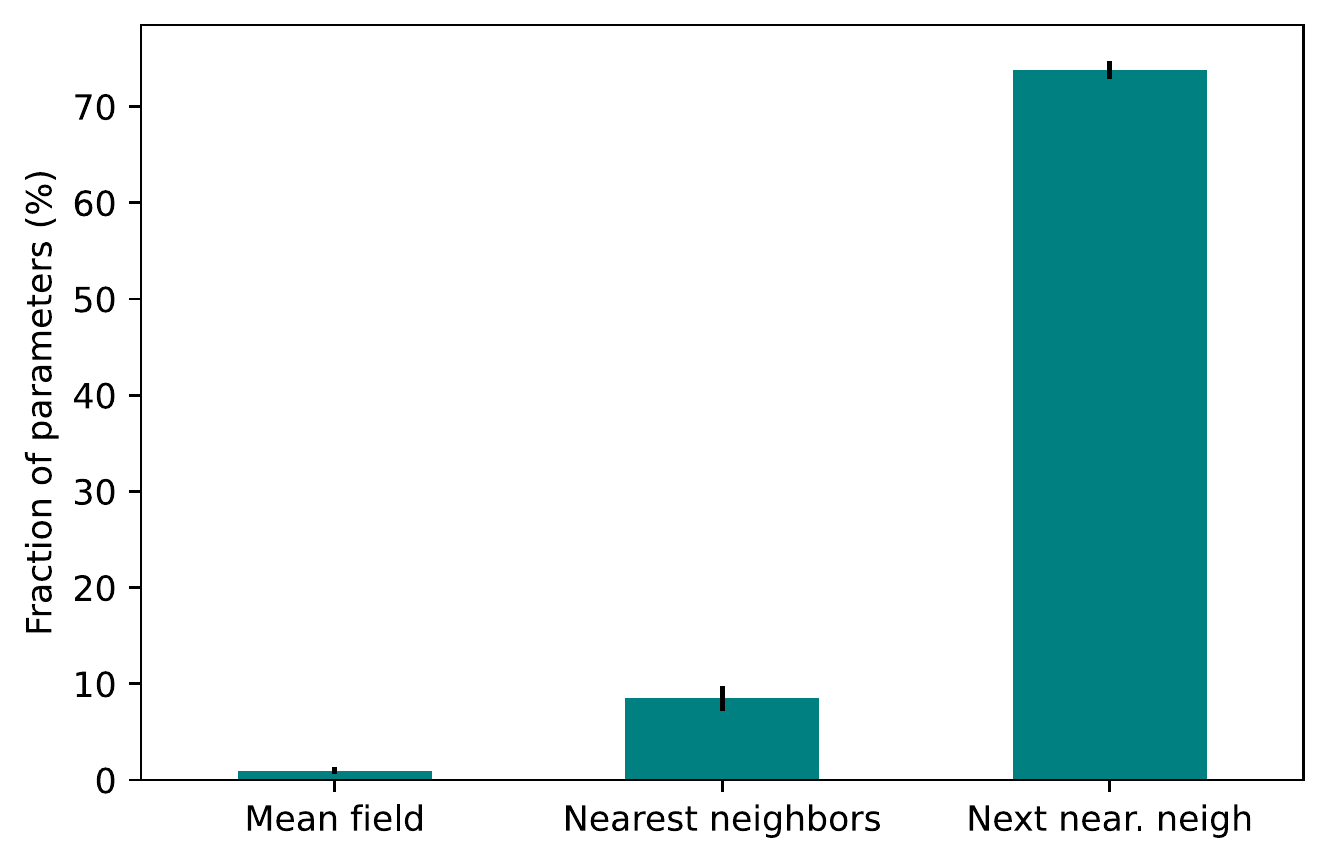}
\caption{Percentage of parameters needed, on average, when considering the approximation of the dependencies between individuals, with respect to the case of no approximation as described in section \ref{sec_neigh_approx}. See figure \ref{fig:pzeroaccuANN} for the accuracy and details regarding the epidemic realizations.}
\label{fig:numpars}
\end{figure}
Figure \ref{fig:pzeroaccuANN} shows how the accuracy in finding the patient zero is influenced by the approximation chosen, while in figure \ref{fig:numpars} the reduction of the number of parameters in the network, with respect to the full graph case, is shown.
We see that the \emph{next nearest neighbors} approximation gives estimates which are on par with the full graph case, while employing less parameters, thus reducing the space and time needed for the training.

\subsection{Training procedure}
To train the whole network, we do 10 000 steps of the annealing procedure described in the main text (except for the \textbf{hospital} case where we do 20 000 steps), increasing $\beta$ linearly from 0 to 1. At each step, we generate 10 000 epidemic cascades before updating the weights using the ADAM optimizer \cite{kingma_adam_2017} with learning rate $lr=0.001$.

In the case of the risk inference problem, we also include a prior probability in the model, with strength decreasing linearly with $\beta$.
This prior is computed from the probabilities of each individual being sampled as $S$, $I$ or $R$ at the last time instant $t=T$, and it is designed in such a way that, when $\beta = 0$, the probabilities for the three states are equal.

The running time of the ANN algorithm on a single instance ranges from one hour to two days on a single GPU (Nvidia TITAN RTX), depending on the structure of the observations $\mathcal{O}$ and that of the underlying contact network. Our implementation using the PyTorch framework \cite{NEURIPS2019_9015} is publicly available in the repository \cite{ann_github}. The results presented in the main text and in SI can be reproduced using the following repository \cite{ann_results_github}.
%Since the \emph{patient zero} problems imply that usually some nodes are observed as $S$ at the final time instant, and there is at least one observation for each individual, the number of configurations that are possible is usually smaller than in the \emph{risk inference} case, and the ANN take less time to train.

\section{Simplified Conditional probability on acyclic graph}\label{sec_acyclic_graph}

In this section we show that in the case of epidemic spreading in a acyclic interaction networks, it is possible to restrict the dependence of previous nodes in the conditional probabilities Eq.7, to the second neighbors with lower index. 
Let us consider the configuration $x$ of $N$ variables, a probability distribution $p(\underline{x})=\frac{1}{Z}\prod_{a}\psi_{a}\left(x_{a}\right)$ factorized over a set of factors $\{a\}$.  We first demonstrate the following statements: 
\begin{lemma}[Markov blanket]
Let  $p\left(x\right)=\frac{1}{Z}\prod_{a}\psi_{a}\left(x_{a}\right)$, and $I\cup J\cup K=\left\{ 1,\dots,N\right\} $ be disjoint and assume no factor depends on $x_{I}$ and $x_{K}$ simultaneously. Then $p(x_I,x_K|x_J)=p(x_I|x_J)p(x_K|x_J)$.\label{lem:blanket}
\end{lemma}

\proof
Considering proportionality $\propto$ with respect to $x_{I}$ only $p\left(x_{I}|x_{J},x_{K}\right)  \propto p\left(x_{I},x_{J},x_{K}\right) \propto\prod_{a}\psi_{a}\left(x_{a}\right) \propto\prod_{a\in\partial I}\psi_{a}\left(x_{a}\right)$ and $p\left(x_{I}|x_{J}\right)  \propto p\left(x_{I},x_{J}\right)\propto\sum_{x_{K}}\prod_{a}\psi_{a}\left(x_{a}\right) \propto\prod_{a\in\partial I}\psi_{a}\left(x_{a}\right)$.As both distributions are normalized wrt $x_{I}$, they must be equal.
This implies that 
\begin{align*}
p\left(x_{I},x_{K}|x_{J}\right) & =p\left(x_{I},x_{J},x_{K}\right)p\left(x_{J}\right)^{-1}\\
 & =p\left(x_{I}|x_{J},x_{K}\right)p\left(x_{J},x_{K}\right)p\left(x_{J}\right)^{-1}\\
 & =p\left(x_{I}|x_{J}\right)p\left(x_{K}|x_{J}\right)
\end{align*}
where the last line follows from the derivation above.

\begin{lemma}[Separated neighborhood]
Let  $p\left(x\right)=\frac{1}{Z}\prod_{a}\psi_{a}\left(x_{a}\right)$, and $G=(V\cup A, E)$ be the associated bipartite factor graph, with $E=\{(i,a)\in V\times A: \psi_a \text{ depends on variable } x_i\}$. Let $I\cup J\cup K\subseteq\left\{ 1,\dots,N\right\} $ be disjoint and assume that every path from $I$ to $K$ in $G$ must pass through $J$ (equivalently, removing vertices in $J$ leave $I$ and $K$ separated).  Then $p(x_I,x_K|x_J)=p(x_I|x_J)p(x_K|x_J)$, and $p(x_I|x_{J\cup K})=p(x_I|x_J)$
\end{lemma}
\proof
Let $I'$ be the connected component of $I$ in $G\setminus J$ and $K'=V\setminus (I'\cup J)$. As all paths from $I$ to $K$ pass through $J$, no factors can depend on variables both in $I'$ and in $K'$. By Lemma \ref{lem:blanket}, $p(x_{I'},x_{K'}|x_J)=p(x_{I'}|x_J)p(x_{K'}|x_J)$. Then $p(x_I,x_K|x_J)=\sum_{x_{I'\setminus I}}\sum_{x_{K'\setminus K}}p(x_{I'},x_{K'}|x_J)=\sum_{x_{I'\setminus I}}p(x_{I'}|x_J)\sum_{x_{K'\setminus K}}p(x_{K'}|x_J)=p(x_I|x_J)p(x_K|x_J)$. Then $p(x_I|x_K,x_J)=p(x_I,x_K|x_J)p(x_K|x_J)^{-1}=p(x_I|x_J)$.

\begin{figure}[h]
\centering
\includegraphics[width=1\textwidth]{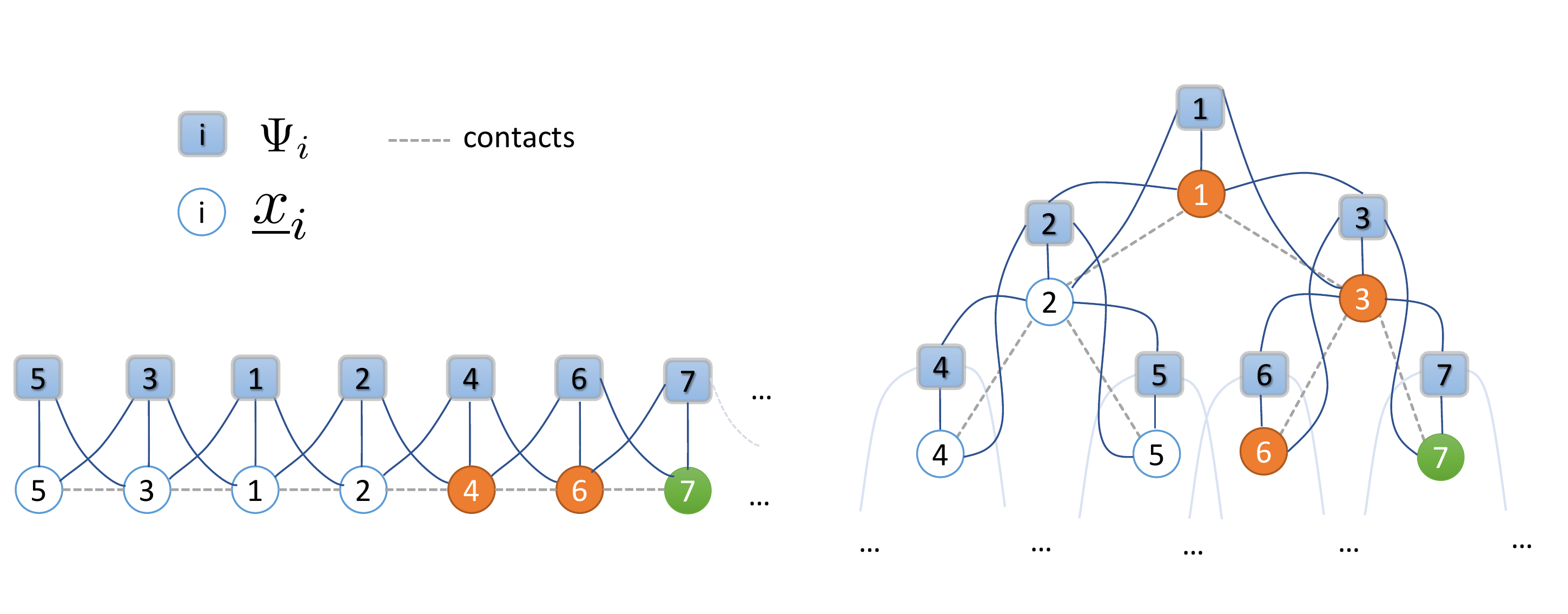}
\caption{Two example of acyclic contact graphs ordered topologically. The two cases represent a linear chain on the left, and a tree on the right. The square represents the factor $\Psi_i$ of the posterior probability of the dynamical process we want approximate. The circle represents the time trajectory of the states $\underline{x_i}$ of each individuals. The dashed lines are the contacts among individuals, and in both cases they generate a acyclic \emph{contact} graph. Instead For instance the nodes $7$ is separated (in the \emph{factor} graph) from the rest of graph containing the nodes with minor index by the orange nodes.}
\label{fig:factor_graph}
\end{figure}

\begin{cor}[Restricted autoregression]
By calling $I=\{i\}$, $J=\{j<i:j\in\partial i\}$ and $K=\{j<i:j\notin \partial i\}$, we obtain that for an ordering of nodes such that $J$ separates $i$ from $K$ we get $p(x_i|\{x_j : j\in\partial i, j < i\})=p(x_i|\{x_j : j < i\})$.
\end{cor}
The last corollary is defined for a single node $i$. In considering the problem of approximating the posterior probability of an epidemic spreading process we distinguish two graphs: the first one, the \emph{contact} graph, the nodes are the time trajectory of the states $\underline{x_i}$ of each individuals and the edges are the contacts between them. The second, the \emph{factor} graph,  has as nodes, again, the time trajectory of individuals and as factors those ($\{\Psi_i\}$) in the equation 6 in the main text. If the \emph{contact} graph is acyclic and the nodes are topological ordered\cite{knuth1997} then, thanks to the corollary 1, the following identity holds:
\begin{equation}
    P(\mathbf{\underline{x}}|\mathcal{O}) = \prod_i P(\underline{x}_i|\mathbf{x}_{<i})
    = \prod_i P(\underline{x}_i|\mathbf{\underline{x}}_{\partial^2 <i})
    \label{eq:nn_cond_prob}
\end{equation}
where $\mathbf{x}_{\partial^2 <i}$ define the sets of nodes up to the next nearest neighbors (in the \emph{contacts} graph) with index lower than $i$. The set of nodes $\mathbf{x}_{\partial^2 <i}$ in the \emph{contact} graph correspond to the $\mathbf{x}_{\partial <i}$ in the \emph{factor} graph. To better visualize the two graphs, in Fig.\ref{fig:factor_graph} we show two cases of \emph{contact} acyclic graphs. The first one represents a linear chain defined by the contacts among individuals. The corresponding \emph{contact} graph is acyclic but the \emph{factor} graph contains cycles. In this case, for instance, considering the conditional probability of node $7$ we have that $P(\underline{x}_7|\mathbf{\underline{x}}_{<i}) =P(\underline{x}_7|\mathbf{\underline{x}}_{\partial^2 <i})=P(\underline{x}_7|\underline{x}_4,\underline{x}_6)$. The node $4,6$ separated the nodes $7$ from the previous nodes in the \emph{factor} graph, as the corollary 1 require. Similarly, the same concept apply to the second case, where we consider a tree graph. In both cases the nodes are ordered topologically so the equation \eqref{eq:nn_cond_prob} holds for each conditional probability.

\section{Robustness of results with respect to the epidemic model}
\begin{figure}[h]
%\centering
\includegraphics[width=1.0\textwidth]{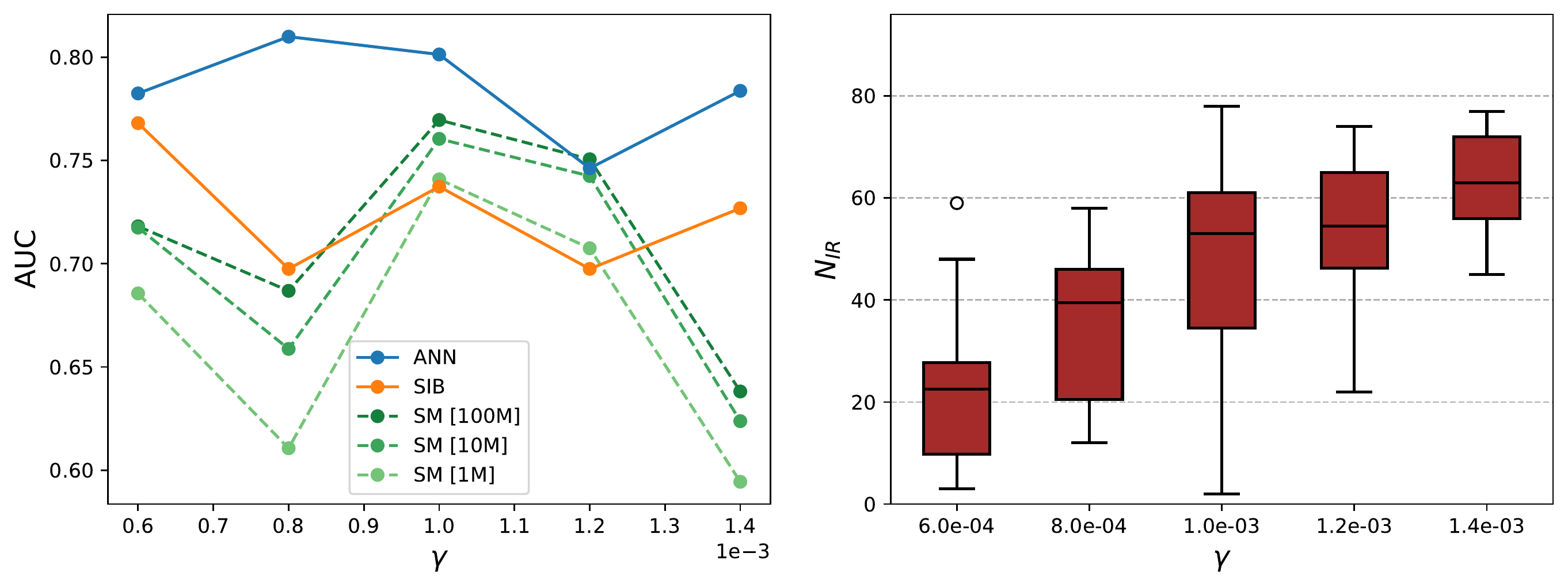}
\caption{Accuracy in finding the patient zero while varying infectivity on the \textbf{work} contact graph. Here we look at the performance by the area under the curve (AUC) of the fraction of patient zeros found as a function of the fraction of considered individuals. We consider 20 different epidemic realizations for the  $\gamma$ values $\gamma=0.6\cdot 10^{-3}$, $\gamma=0.8\cdot 10^{-3}$, $\gamma=1.2\cdot 10^{-3}$, $\gamma=1.4\cdot 10^{-3}$,  and 100 realizations for  $\gamma=10^{-3}$, while the recovery rate is fixed $\mu=0.02$. On the right pane, the boxplot shows the number of infected individuals with the different values of $\gamma$, in a total population of 95 individuals. }
\label{fig:auc_gamma}
\end{figure}

The main text shows results for different classes of epidemic inference problems. These problems depend on several parameters that could change the scenario analyzed. For instance, in the case of patient zero problems, we can vary the time $T$ of the observed snapshot, the number of individuals, or the degree of the graphs in the case of synthetic interaction networks or the epidemic like the infectious or recovery parameters. All cases analyzed show strong fluctuations, imposing to average over a large number of instances to have a proper statistical significance of the results and appreciate the difference in the performance of the methods compared. As specified in the main text, we chose the infectious and recover parameters from having, on average, half of the infected individuals to decrease the instances where very few or almost all of them are infected (see fig.\ref{fig:auc_gamma}, right plot). In these cases, the inference problems analyzed become trivial or impossible to solve. On the other hand, our method has a computational cost that spans several hours on modern GPU to reach convergence for each instance analyzed, limiting the possibility of exploring the large dimensional space of the parameters of our inference problems. Nevertheless, we check how the performances of the methods analyzed vary, in the patient zero and the inference of parameters problems, with respect to the infectious parameters ($\lambda$ or $\gamma$) to check the robustness of the results shown in the main text.

%Here we investigate the robustness of our method with respect to changing the parameters of the epidemic model. In particular, we focus on the infectivity parameter ($\lambda$ or $\gamma$) in order to explore different epidemic scenarios.

For the patient zero inference problem, we consider the \textbf{work} contact graph, and look at the performance of the different algorithms used, changing the infectivity $\gamma$. The figure \ref{fig:auc_gamma} shows the area under the curves representing the fraction of times the patient zero is found, as in Figure 4 of the main text (right plots). The ANN method shows consistent performance with changing $\gamma$, giving the best score or on par with others algorithms. 

\begin{figure}[h]
\centering
\includegraphics[width=0.8\textwidth]{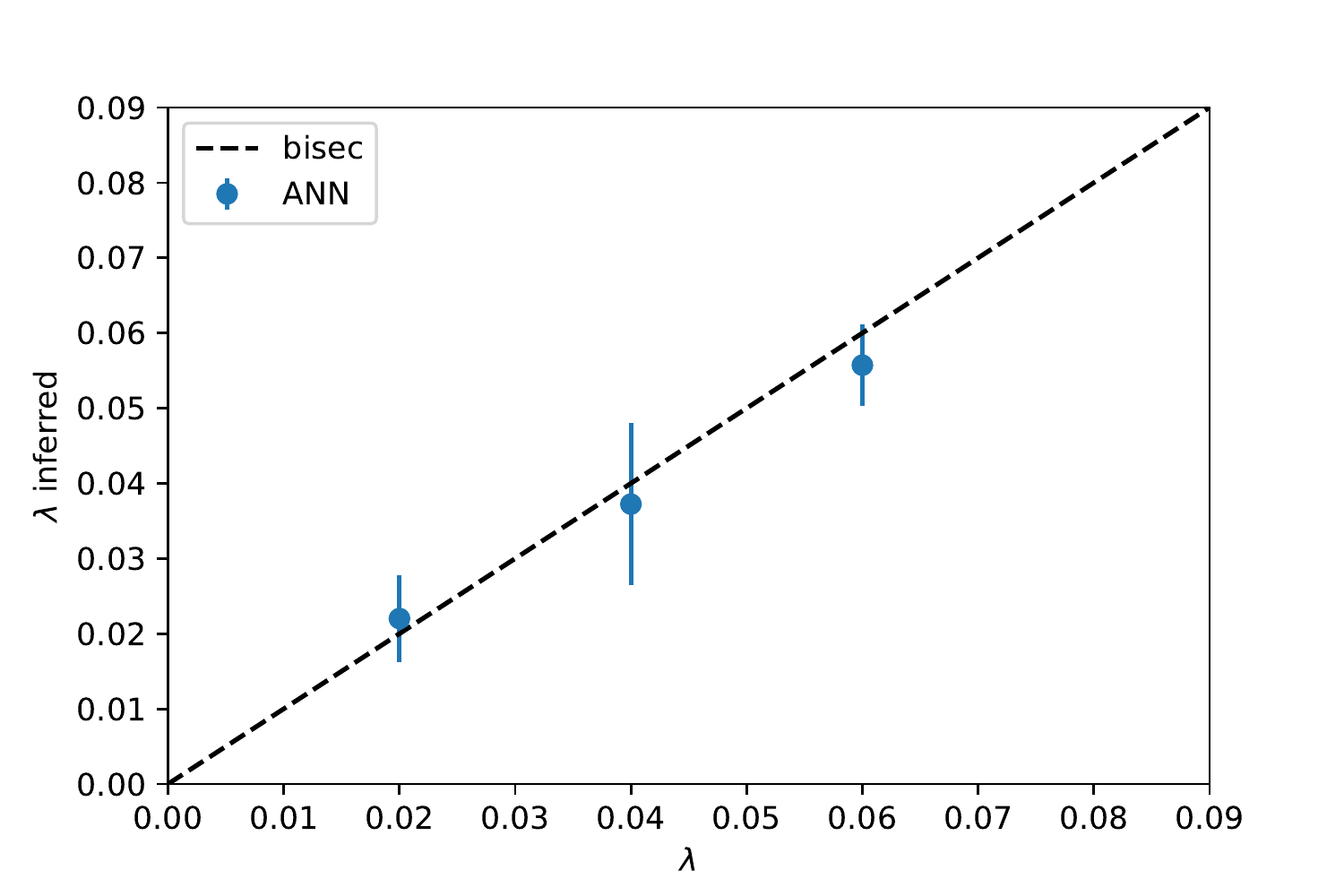}
\caption{Inference of infectiousness parameter $\lambda$ in the {\bf rrg} case (100 individuals interacting according a random regular graph with degree $10$ for 15 days). The problem setting is equals to that explained in the subsection "Epidemic Parameters inference" in the main text. The plot shows the infectious parameter $\lambda$ inferred by ANN with the perfect inference line $\lambda_{est} = \lambda$ highlighted. ANN correctly estimated them in error margin.}
\label{fig:infer_lam}
\end{figure}

Then we check the robustness of the inference of parameters in the case of of \textbf{rrg} interaction graph; the figure \ref{fig:infer_lam} shows that the inference of the infectivity parameter $\lambda$ remains good as $\lambda$ varies.

\section{Soft margin estimator\label{sec:softm}}

We also use the Soft Margin estimator from \cite{antu_identification_2015} for finding the patient zero. This is a Monte Carlo based estimator, which applies Bayes formula to estimate the source probability $\mathbbm{P}\left(s=i \right)$: given an epidemic $\mathbf{x}(i)$ which is the result of the simulation, starting from individual $i$, and the set of observation $\mathbf{x}_O$ on the epidemic, the probability of node $k$ being the source of the epidemic is
\begin{equation}
    \mathbbm{P}\left(s=k  \mid \mathbf{x}_O\right)  \propto P\left(\mathbf{x}_O \mid s=k\right) P\left(i\right)
\end{equation}
We have extended the method to the SIR model.
To evaluate $P\left(\mathbf{x}_O \mid s=k\right)$, we run $M$ Monte Carlo simulations in which the source of the epidemic is $k$ and compare the resulting epidemic cascade $\mathbf{x}_i$ with the observations. 
For this we use the Jaccard similarity function $\phi_X \left(\mathbf{x}_i, \mathbf{x}_O\right)$, relating how many individuals are in state $X$ (either infected, $X=I$ or recovered, $X=R$) in the generated cascade and the observations. 
Clearly, if no individuals are observed in state $X$, then $\phi_X = 1$ regardless of the realization.
The probability of observing a certain configuration, given the source of the epidemic, can then be written as:
\begin{equation}
    P\left(\mathbf{x}_O \mid s=k\right) = \frac{1}{M} \sum^M_{i=1} 
    \exp{\left\{ - \frac{\left( 1-\phi_I \left(\mathbf{x}_i, \mathbf{x}_O\right)\right)^2
    +\left( 1-\phi_R \left(\mathbf{x}_i, \mathbf{x}_O\right)\right)^2
    }{a^2}\right\}}
\end{equation}
where the coefficient $a$ regulates the sharpness of the peak around the case of perfect matching of observations ($\phi_I = 1$ and $\phi_R = 1$), where the corresponding value in the sum is $1$. For the patient zero problems, we select the alpha values that give better results (retrospectively) because they seem to be strongly dependent on the contact graphs and the number of samples considered.
%%In particular for $a\rightarrow 0$ we should get in the regime of "classical" Monte Carlo, effectively counting the number of simulations that match the observations.
\\
%We use the Soft Margin estimator as a benchmark for our method, using an efficient implementation of the estimator for the simulations.

\bibliographystyle{unsrtnat}
\bibliography{references}